\PassOptionsToPackage{dvipsnames,svgnames}{xcolor}

\documentclass[sigconf]{acmart}

\copyrightyear{\copyright{} Zeyu Huang, Xinyi Cao, Yuanhao Zhang, Xiaojuan Ma, 2024. This is the author's version of the work. It is posted here for your personal use. Not for redistribution. The definitive Version of Record was published in ACM Digital Library, http://dx.doi.org/10.1145/3613904.3642818}
\acmYear{2024}
\setcopyright{none}
\acmConference[CHI '24]{Proceedings of the CHI Conference on Human Factors in Computing Systems}{May 11--16, 2024}{Honolulu, HI, USA}
\acmBooktitle{Proceedings of the CHI Conference on Human Factors in Computing Systems (CHI '24), May 11--16, 2024, Honolulu, HI, USA}
\acmDOI{10.1145/3613904.3642818}
\acmISBN{979-8-4007-0330-0/24/05} 

\usepackage[shortlabels]{enumitem}
\usepackage{xspace}
\usepackage{multirow} 
\usepackage{subcaption} 
\usepackage{tabularx}

\newcommand{\edit}[1]{{\color{violet}#1}}
\renewcommand{\edit}[1]{#1}

\newcommand{\hkust}{
  \affiliation{%
    \institution{The Hong Kong University of Science and Technology}
    \city{Kowloon}
    \country{Hong Kong S.A.R.}
  }
}
\newcommand{\atconnect}{@connect.ust.hk}

\begin{document}
\title[Sharing Frissons among Online Video Viewers]{Sharing Frissons among Online Video Viewers: Exploring the Design of Affective Communication for Aesthetic Chills}


\author{Zeyu Huang}
\email{zhuangbi\atconnect}
\orcid{0000-0001-8199-071X}
\hkust{}

\author{Xinyi Cao}
\authornote{This research project is conducted during the author's visiting at The Hong Kong University of Science and Technology}
\email{xcao86@gatech.edu}
\orcid{0009-0005-8601-5039}
\affiliation{%
  \institution{Georgia Institute of Technology}
  \streetaddress{North Avenue}
  \city{Atlanta}
  \state{Georgia}
  \country{USA}
  \postcode{30332}}

\author{Yuanhao Zhang}
\email{yzhangiy\atconnect}
\orcid{0000-0001-8263-1823}
\hkust{}

\author{Xiaojuan Ma}
\authornote{Corresponding author}
\email{mxj@cse.ust.hk}
\orcid{0000-0002-9847-7784}
\hkust{}

\renewcommand{\shortauthors}{Huang et al.}

\begin{abstract}
  On online video platforms, viewers often lack a channel to sense others' and express their affective state on the fly compared to co-located group-viewing. This study explored the design of complementary affective communication specifically for effortless, spontaneous sharing of frissons during video watching. Also known as aesthetic chills, frissons are instant psycho-physiological reactions like goosebumps and shivers to arousing stimuli. We proposed an approach that unobtrusively detects viewers' frissons using skin electrodermal activity sensors and presents the aggregated data alongside online videos. Following a design process of brainstorming, focus group interview (N=7), and design iterations, we proposed three different designs to encode viewers' frisson experiences, namely, ambient light, icon, and vibration. A mixed-methods within-subject study (N=48) suggested that our approach offers a non-intrusive and efficient way to share viewers' frisson moments, increases the social presence of others as if watching together, and can create affective contagion among viewers.
\end{abstract}



\begin{teaserfigure}
  \centering
  \includegraphics[width=.9\textwidth]{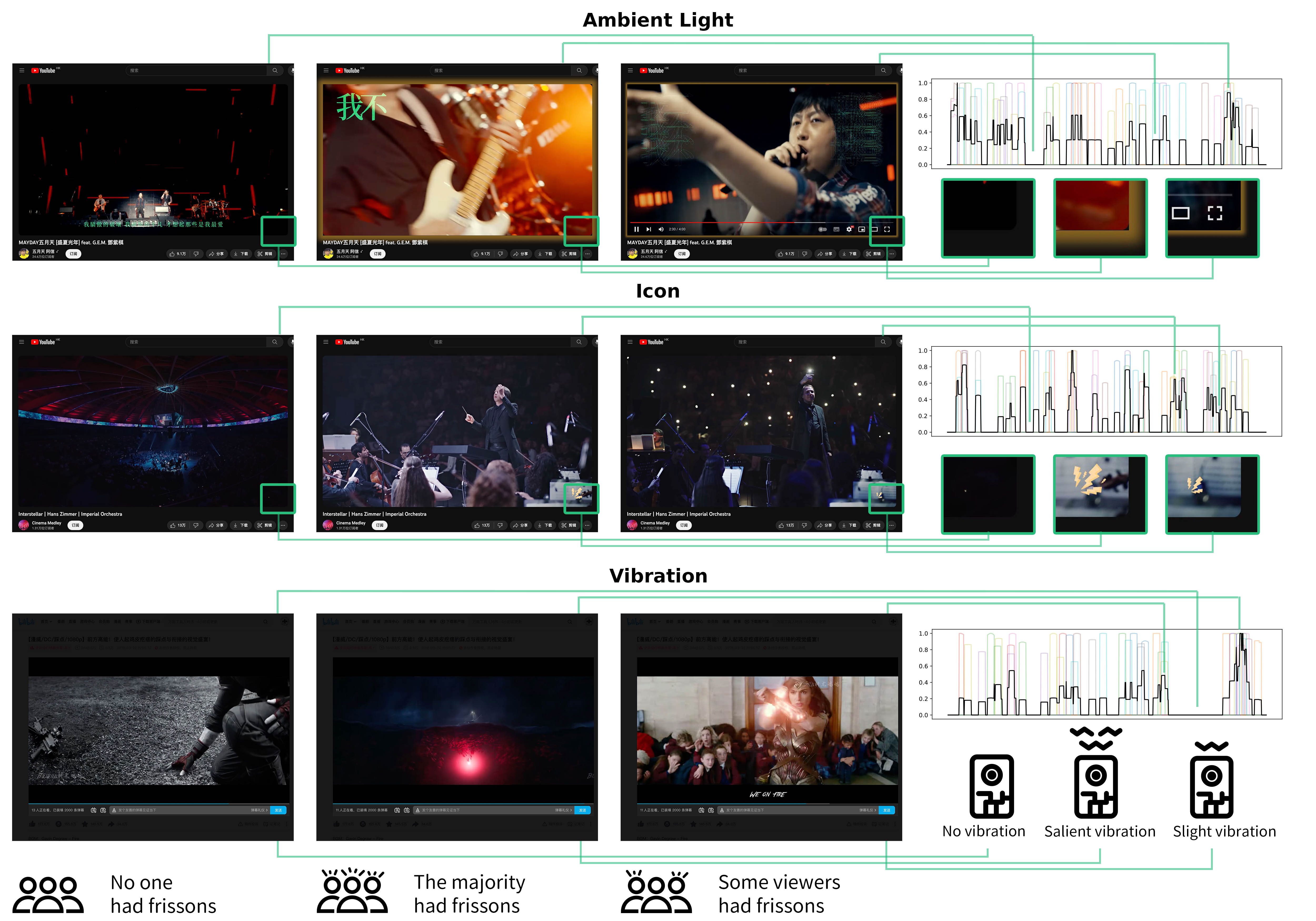}
  \caption{Three different designs to present the viewers' collective frisson experiences.
  Each row shows one design applied to one online video,
  and it contains three moments when no user
  experiences frissons, most users experience frissons, and some users experience frissons.
  One the right of each row is a plot of the viewers' frisson data while watching the video.
  \edit{The bolt icon is designed by Freepik.}}
  \Description{Three rows of screenshots from three online videos. The first column of each row is peaceful moments of videos where no significant additional feedback is present. The second column is climax moments, where we can also see a large ambient halo around the video, a large lightning icon at the corner, or a salient vibration. The third column is moderately exciting moments, where we can see a small ambient halo around the video, a small lightning icon at the corner, or a moderate vibration.}
  \label{fig:teaser}
\end{teaserfigure}


\maketitle

\section{Introduction}

Frisson is a psycho-physiological phenomenon commonly perceived by people while watching emotionally arousing works such as music performances and movies~\cite{goldsteinThrillsResponseMusic1980}.
It is also referred to as chills, aesthetic chills, or thrills,
where people feel a short but intense tingling sensation accompanied by
goosebumps or shivers down the spine~\cite{bannisterDistinctVarietiesAesthetic2019,goldsteinThrillsResponseMusic1980}.
Such feelings are tightly linked to the moments of pleasure in the brain~\cite{sachsBrainConnectivityReflects2016}
and strong emotional experiences~\cite{salimpoorRewardingAspectsMusic2009}
while appreciating rewarding stimuli. 
Apart from a stimulus itself,
another account of frisson is shared experiences~\cite{bannisterDistinctVarietiesAesthetic2019}.
Together in an audience, 
people consciously or unconsciously sense one another's signs of physiological and psychological states, e.g., heartbeats and warmth~\cite{inagakiSharedNeuralMechanisms2013}, which consequently transfers moods and emotions among individuals~\cite{gernotEmpathyEinfuhlungAesthetic2018}. 
Frisson can also be a product of such affective contagion. 
That is to say, the frisson experience can turn into a social process when people occasionally want to share their goosebump moments with others~\cite{neidlingerAWElectricThatGave2017,heFrissonWavesExploring2022}.

We noticed that online video platforms (e.g., YouTube) contain abundant content
of music performances and cinematic videos ---
common stimuli of frisson.
The viewers of these videos also mention their frisson experiences from time to time
in the comment area~\cite{thelwallCommentingYouTubeVideos2012}
or using time-synced comments (a.k.a.\@ danmaku) overlaid on the video~\cite{maVideobasedEvanescentAnonymous2017}
to accompany their emotional expression of being
stunned, touched, in awe, etc.
However,
users' opportunities to sense the thrills of other viewers or pass their thrills on to other viewers are limited
due to the distributed, asynchronous nature of such platforms.
Both regular comments and danmaku verbally communicate with people watching the same video at different places and times.
They largely fail to capture the passing moment of frisson in a timely manner, as the former is detached from the video timeline and the latter often entails a delay~\cite{maVideobasedEvanescentAnonymous2017}.
Besides, posting comments requires extra effort to externalize the sudden feelings, which may distract individuals from their video-viewing experiences.

To mitigate this issue,
we propose to explore the design of an augmentation to existing online video platforms
that can facilitate the sensing and sharing of frisson in a lightweight yet effective manner.
Such a communication channel will not replace any current text-based interactions; rather, it will serve as a supplement and optional feature focusing on the timely communication of group
psycho-physiological reactions at the thrill-arousing moments.
Our goal is to create a sense of ``watching together'' for online video viewers and amplify the affective contagion among the audience through frisson sharing~\cite{wetherellAffectEmotionNew2012,guoPreliminaryStudyEmotional2020}.
While it may not be suitable for every video and needed by every viewer,
we foresee its particular benefit for those people desiring to share their
feelings while watching videos with other viewers,
and its especial compatibility with those artistically captivating videos.

Prior studies showed the efficacy of using biosignals as an effective and unobtrusive channel for affective communication in contexts such as
face-to-face chitchatting~\cite{rojasProjectUsWearable2020},
instant messaging~\cite{hassibHeartChatHeartRate2017},
and maintaining long-distance relationship~\cite{wernerUnitedpulseFeelingYour2008}.
We adopt a similar approach because changes in biosignals like skin conductance (EDA, a.k.a.\@ GSR)
are considered a major objectively measurable physiological response of frisson~\cite{wassiliwizkyEmotionalPowerPoetry2017,bannisterDistinctVarietiesAesthetic2019,heFrissonWavesExploring2022,neidlingerAWElectricThatGave2017}.
Wearable sensors can non-intrusively capture such variations in biosignals and help infer people's corresponding affective states during an aesthetic experience~\cite{tschacherPhysiologicalSynchronyAudiences2023}. 
We use these sensing data as design materials for affective communication, investigating different representations of collective frisson signals that can create an engaging shared viewing experience while maintaining users' primary attention on the video content. 

\begin{itemize}
    \item[\bfseries RQ1] Will sensing and feedback of viewers' biosignals facilitate effective communication of frisson moments on online video platforms?\label{rq1}
    \item[\bfseries RQ2] Will sensing and collective sharing of viewers' frisson signals improve a sense of group affective experiences on asynchronous online video platforms?\label{rq2}
    \item[\bfseries RQ3] What are the design considerations for collective affective biosignal feedback on online video platforms to provide the best user experience?\label{rq3}
\end{itemize}

To this end,
we propose a novel asynchronous affective communication approach among viewers of the same online videos.
As is shown in Figure~\ref{fig:teaser},
it uses EDA sensors to non-obtrusively capture viewers' frissons while watching videos,
and provides peripheral feedback of the audience's collective frisson data to viewers.
From a thorough design process of brainstorming, focus group interview (N=7), and further iterations,
we derive three different feedback designs to convey viewers' frisson data
with glowing ambient light, a size-changing icon, and a fingertip linear vibration motor.
In a mix-methods, within-subject study with 48 participants,
we evaluated users' overall attitudes towards such a frisson-sharing channel on online video platforms,
and we compared the efficiency of the three designs of frisson feedback.
The results show that such a frisson-sharing approach can efficiently ease the sender's cognitive load
of initiating affective communication while watching online videos,
and it can deliver other viewers' frisson experiences to the receiver in a lightweight and straightforward way.
Participants also suggested the strong social attribute of this affective communication channel.
Meanwhile,
we noted participants' different habits of viewing online videos,
especially their dissimilar needs for social connections with other viewers.
Such a difference also leads to their diverse attitudes towards this communication channel.

The key contributions of this research are as follows:
\begin{itemize}
    \item An affective communication channel for online video platforms to gather viewers' frissons unobtrusively and provide collective feedback with minimal cognitive loads.
    \item A mix-method evaluation including users' overall attitudes and a comparison between various design alternatives.
    \item A summary of design considerations for future work on asynchronous affective communication systems using biosignals.
\end{itemize}

\edit{Our findings demonstrate the feasibility of using physiological reactions like frissons as ``emotional prosthesis'' with the help of ever-developing sensing technologies.
We also anticipate a wider usage of sensor-detected frissons as an affective communication channel beyond the scenario of online videos.
Researchers and designers can seek inspiration from our discussion of participants'
perceptions of the social values behind the shared physiological data,
as well as their attitudes towards sharing frissons in an asynchronous and auxiliary manner.}
\section{Related Work}

\subsection{Feedback of Audience's Activities}

It has already been discovered that when people watch sports events, music events, etc., in groups or crowds,
they often have a better affective experience than watching alone~\cite{guoPreliminaryStudyEmotional2020,garridoAudienceEngagementCommunity2020,tschacherPhysiologicalSynchronyAudiences2023},
as the interpersonal spontaneous or serendipitous interactions often lead to emotional resonance~\cite{herrandoEmotionalContagionBrief2021}.
Such a phenomenon drives the exploration of designs that strive to provide a similar atmosphere for remote audiences who lack the social cues of other viewers.
For example, Vatavu designed Audience Silhouettes, a visualization of other audience's body movements below the TV screen when they watch the same sports event on TV~\cite{vatavuAudienceSilhouettesPeripheral2015}.
Empatheticons by Chen et al.\@ animates users' avatars on an online music platform
to provide metaphors for their emotional reactions~\cite{chenEmpatheticonsDesigningEmotion2014}.
Such works simulated the non-verbal communication and the overall atmosphere of watching/listening in a crowd.
The trending live-streaming services display intensive real-time data streams of activities among concurrent but distributed viewers, 
including real-time comments, likes, and gifts~\cite{luYouWatchYou2018}.

However, the aforementioned designs are mainly for synchronous viewing scenarios.
On regular online video platforms such as YouTube, audience activities around a video are often not simultaneous in time and thus are seldom presented or utilized to directly stimulate audience members' affective responses on the fly.
Traditionally, users of these platforms may reveal their thoughts and feelings by leaving comments beneath a video, which mainly allows post-hoc textual discussions in a threaded manner.
It is hard to convey viewers' instant experience of emotion through this channel as it requires significant attention and context switching between the video window and the comment box.
Recently, time-synced comments (originated from \emph{Danmaku}) have been introduced into mainstream online video platforms.
Short comments (often text-based) posted non-concurrently appear on the videos based on their anchors along the video timeline, creating a sense of watching together~\cite{chenWatchingMovieAlone2017}. 
While time-synced comments enable audiences to show their reactions towards a particular moment in the video, it demands extra cognitive efforts to produce and/or comprehend textual information while watching the video~\cite{chenWatchingMovieAlone2017}. 
This may distract users and interrupt their viewing experiences~\cite{maVideobasedEvanescentAnonymous2017}.
Moreover, such commentaries often incur a delay behind the actual critical moments in the video due to the time needed to type them out.
When Danmaku is displayed on top of videos and obscures the original content,
many users find the feature annoying and end up turning it off~\cite{maVideobasedEvanescentAnonymous2017}.

Concerning the cognitive demand for processing comment-based communication among remote audiences,
new designs have emerged to represent accumulative viewer behaviors so as to enhance the group-watching experience on online video platforms.
For instance, on YouTube, a recent feature called ``popular segments'' shows the replay frequency of each portion of the video by a line plot above the progress bar.
Such a feature mainly provides informational assistance for users to locate the most essential parts of a video based on combined actions. 
A similar graph above the time bar is employed in a Chinese video platform, Bilibili\footnote{\url{https://www.bilibili.com/}},
which, instead, reveals \emph{Danmaku} density at each moment. 
It can hint the moments that draw significant attention from the viewers even when a user hides the \emph{Danmaku} comments. 
These examples showcase the feasibility of using simple graphics to convey collective responses to different moments in a video; but still, they require intentional user actions as the input data. 
To reduce the data production effort on the user end,
our work proposes another complementary affective communication approach for online video viewers using biosignal sensors,
as they can automatically record the data related to viewers' affective experience.
Moreover, besides senders' lower cognitive efforts,
researchers have discovered that
sharing biosignal data can effectively foster affective communication
in several other use scenarios.
More on this is discussed in the following subsection.

\subsection{Biosignal Sharing for Affective Communication}\label{sec:rw:biosignal}

Biosignals are body variables that can be measured and monitored~\cite{feijtSharingBiosignalsAnalysis2021} 
Common biosignals include heart rates (HR),
heart rate variability (HRV),
and skin electrodermal activity (EDA), to name a few.
They are highly related to humans' psychological activities~\cite{betellaInferenceHumanAffective2014}.
For example,
frisson experiences can be detected from changes in EDA~\cite{craigExploratoryStudyPhysiological2005}.
Sharing biosignal data or their derived information
can facilitate affective communication between either
co-located~\cite{rojasProjectUsWearable2020,dsouzaAugmentingCoLocatedSocial2018,semertzidisNeoNoumenaAugmentingEmotion2020}
or remote users~\cite{hassibHeartChatHeartRate2017,wernerUnitedpulseFeelingYour2008,kimBreathingFrameInflatableFrame2015}.
For one thing, with the help of well-developed biosignal sensors, such data can be captured and sent in real time with minimal attention or cognitive load imposed on users~\cite{semertzidisNeoNoumenaAugmentingEmotion2020,minBiosignalSharingAffective2014}. 
For another, raw biosignals and inferred affective states can be expressed through diverse modalities. 
Visual feedback is the most common way to display individual as well as collective biosignal-related information. It has an ample design space ranging from line plots that focus on accuracy and intuitiveness to abstract patterns that better meet aesthetical needs~\cite{liuCanBiosignalsBe2017,snyderMoodLightExploringPersonal2015,qinHavingHeartTime2021}. 
Previous works also explore the use of haptic (e.g., vibration) or auditory modalities to simulate the authentic physical traits of certain biosignals~\cite{howellLifeAffirmingBiosensingPublic2019,minBiosignalSharingAffective2014}. 

The ease of sensing and richness of representation empowers many applications of biosignals in interpersonal affective communication. 
For example,
NeoNoumena enhanced face-to-face interactions
by inferring users' emotions from ECG data and visualizing them in abstract patterns in AR around them~\cite{semertzidisNeoNoumenaAugmentingEmotion2020}.
For remote communication, HeartChat extends instant messaging apps for users to see each other's heart rates in line charts and colored message bubbles~\cite{hassibHeartChatHeartRate2017}. 
Liu et al.\@ made a smartwatch application to prompt active sharing of users' heart rates at emotional moments with their contacts with pre-compiled template messages~\cite{liuSupportingSocialInteractions2017}.
Researchers also designed tangible objects as mediums of biosignal sharing among physically separated individuals who are in close relationships,
such as vibrating rings~\cite{wernerUnitedpulseFeelingYour2008} and inflating photo frames~\cite{kimBreathingFrameInflatableFrame2015}
to create a sense of physical togetherness.
Participants in these research studies confirmed that
sharing biosignals is a viable way of exchanging affective experiences~\cite{semertzidisNeoNoumenaAugmentingEmotion2020,kimBreathingFrameInflatableFrame2015,liuSupportingSocialInteractions2017,hassibHeartChatHeartRate2017}, which can psychologically draw users closer to one another~\cite{wernerUnitedpulseFeelingYour2008,liuSupportingSocialInteractions2017,hassibHeartChatHeartRate2017}.

Nevertheless, the above examples exploited biosignal feedback in contexts where communication (synchronous or asynchronous) is the primary task. In other words, this body of research is about deliberate affective communication. 
Some researchers investigated the alternative use of biosignals to hint the emotional state of individuals while they perform other tasks together, which was shown to improve collaboration~\cite{tanInvestigatingEffectsUsing2014,deyEffectsSharingRealTime2018}
and engagement level~\cite{walminkDisplayingHeartRate2014}.
This line of research studies auxiliary affective communication that helps maintain a sense of affinity without distracting users from their primary tasks~\cite{deyEffectsSharingRealTime2018}.
Our work falls in this area and focuses on a relatively under-explored context: auxiliary affective communication in asynchronous experiences. 
We are interested in the problem of \textbf{how people receive others' ``spontaneous'' reactions and whether such reactions affect their own when they engage in the same activity at different spaces and times}. 
In the scope of this paper, we use the scenario of online video-watching to experiment with the design of frisson sharing for creating a group viewing experience.

\subsection{Detecting and Sharing Frissons}
Frisson is a common psycho-physiological reaction
of tingling skin, goosebumps, and shivers~\cite{pankseppEmotionalSourcesChills1995}
induced by arousing and aesthetically pleasing stimuli
such as music, movies, literature, natural scenes, etc.~\cite{goldsteinThrillsResponseMusic1980,wassiliwizkyEmotionalPowerPoetry2017}.
In general, it is a strong indicator of people's affective states~\cite{salimpoorRewardingAspectsMusic2009,sachsBrainConnectivityReflects2016};
in particular,
it is a complicated joint result of
sensory stimuli,
aesthetic appraisal,
affective contagion from other viewers,
and so on~\cite{harrisonThrillsChillsFrissons2014}.

The most common approach to detect frissons is through biosignals.
Rises in EDA
(skin electrodermal activity, a.k.a.\ GSR) are found to be
strongly associated with frisson events~\cite{craigExploratoryStudyPhysiological2005,wassiliwizkyEmotionalPowerPoetry2017}.
There are also ways to sense the goosebumps or skin hair erection (i.e., piloerection)
through image analysis~\cite{benedekObjectiveContinuousMeasurement2010,benedekPhysiologicalCorrelatesEmotional2011}
and skin deformation detection~\cite{kimFlexibleSkinPiloerection2014}.
However, these bodily symptoms are largely affected by other factors, e.g., human's thermal comfort.
Thus are not a reliable cue of frisson~\cite{kimBreathingFrameInflatableFrame2015,wassiliwizkyEmotionalPowerPoetry2017}.
In our work, we use EDA signals captured by wearable sensors to detect frisson moments.

Due to the tight association with affective states,
prior research also looked into frisson sharing to communicate affective states among individuals, a usual practice in people's daily life~\cite{pennebakerDisclosingSharingEmotion2001,rimeEmotionElicitsSocial2009}.
For example, Neidlinger et al.\@ designed AWElectric,
smart clothing that animates the lights and inflation on the clothes
and plays subtle tickling sounds
to exchange frissons between dyads~\cite{neidlingerAWElectricThatGave2017}.
They aimed for an artistic expression of such a sensation
as well as increased sharers' awareness of their own affects~\cite{neidlingerAWElectricThatGave2017}.
Another system, Frisson Waves developed by He et al.,
can transmit the frisson experiences among audiences during a classical music concert
~\cite{heFrissonWavesExploring2022}.
To simulate the contagious effect, it used thermo-haptic feedback, which is discovered to be able to evoke synthetic frissons~\cite{jainFrissonLeveragingMetasomatic2022,harrisonThrillsChillsFrissons2014}, to trigger frisson in a wave-like pattern over the crowd. 
Frisson Waves showed that synthetic frissons could
enhance the sense of connectedness and promote physiological synchrony among co-located audience in a field study.
Nevertheless, users may feel confused about the source of artificial frisson episodes because,
unlike naturally occurring frissons, they may not be accompanied by actual affective resonance with a thrill-inducing stimulus (e.g., a particular music segment).

Instead of focusing on the sensory stimuli aspect of frisson~\cite{harrisonThrillsChillsFrissons2014},
synthesizing frisson episodes,
or creating a
physiologically similar experience for the entire audience,
our work focuses more on the affective impact of sharing frisson episodes among viewers on their video-watching experiences
and how to realize efficient communication of such information.
We seek to create an illusion of ``cybernetic togetherness''~\cite{heFrissonWavesExploring2022} among distributed, asynchronous audiences of rewarding stimuli --- online videos.
Even though they experience the same stimulus separately,
this group of users also demands frequent communication of their affective states~\cite{yasminaUsingYouTubeComments2016}.
However, due to the asynchrony of the viewers' activities
and differences in media
compared to previous face-to-face, immersive scenarios,
it is yet to be explored how to aggregate and encode other viewers' frisson information in the feedback.
Following previous works,
we employed a thorough design process and a subsequent user study to explore
effective designs for sharing natural frissons among online video viewers.

\section{Design}

\begin{figure}[h]
  \centering
  \includegraphics[width=\linewidth]{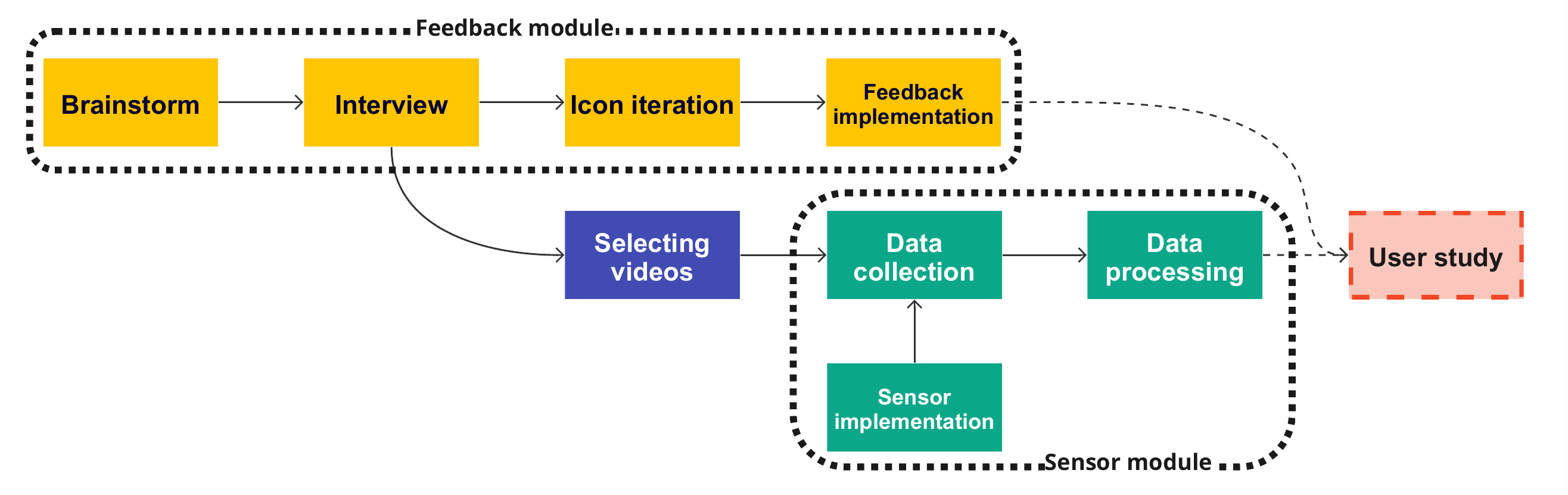}
  \Description[A flowchart describing the order and the dependencies of each subsection.]{The feedback module contains a brainstorm, an interview, an iterative icon design, and the implementation, in that order. The sensor module contains the implementation, the data collection process, and the data processing, in that order. The data collection stage depends on the selection of videos, which further depends on the interview when designing the feedback module. Finally, the processed data and the feedback module are used in the user study.}
  \caption{The design process}\label{fig:design-process}
\end{figure}

This section describes the design and
implementation\footnote{The source code can be found at \url{https://github.com/fhfuih/CHI24-Frissons-Online-Videos}.}
of our proposed frisson-sharing channel.
It consists of two integral parts:
1) a \textbf{sensor module} that captures and stores users' biosignals while watching online videos; and 
2) a \textbf{feedback module} that conveys previous viewers' recorded frisson experiences as the video progresses.
~\autoref{fig:design-process} gives an overview of our design workflow. The rest of this section is organized as follows. 
Section~\ref{sec:design:process} introduces the design process we followed
to arrive at the three final design candidates
for the \textbf{feedback module}.
Then, in Section~\ref{sec:design:video},
we present the steps we undertook to select four sample clips
that represent typical frisson-triggering video genres
based on existing literature and a focus group interview.
Such a selection process aims to trigger a sufficient number of frissons
and assist in the evaluation of the frisson feedback design.
Next,
Section~\ref{sec:design:data} details
the implementation of the \textbf{sensor module}
and the following procedure of
collecting 20 participants' biosignal data while watching the sample videos.
Section~\ref{sec:design:impl} discusses
the implementation of the \textbf{feedback module}.

\subsection{Designing Feedback of Collective Frisson Experiences}\label{sec:design:process}

The design process of frisson feedback comprises three parts:
a brainstorming session, a focus group interview, and an iterative design of plausible candidates.

\subsubsection{Brainstorming of Modalities and Features for Encoding Frisson}
To design appropriate ways to convey viewers' frisson experiences from their biosignal data,
we start by gaining a comprehensive understanding of existing biofeedback designs.
We initially focused on the top HCI venues ranked by Google Scholar~\cite{googlellcHumanComputerInteraction},
using the search query \texttt{``biofeedback OR ((biosignal OR biometric OR physiological OR somatosensory OR <common biosignals' names>) AND (social OR interaction OR interpersonal OR communication))''},
and filtered out works that did not propose new biofeedback designs.
The scope was refined by scanning the citation network of the initially acquired papers
for missing-out work.
We ended up with
37 
papers in total (listed in supplementary material).
Then, we established a digital moodboard of existing biosignal representations presented in these papers to view the common design choices.
In general, among the design examples we extracted,
biosignal information was provided to users in three main modalities:
visual (e.g.,~\cite{semertzidisNeoNoumenaAugmentingEmotion2020,elaliBreatheWithMeExploringVisual2023,kooImmersiveEmotionsTranslating2022,leeUnderstandingDesigningAvatar2022,liuSignificantOtterUnderstanding2021,hassibHeartChatHeartRate2017}),
haptic (e.g.,~\cite{rojasProjectUsWearable2020,wernerUnitedpulseFeelingYour2008,kimBreathingFrameInflatableFrame2015,aslanPiHeartsResonatingExperiences2020}),
and auditory (e.g.,~\cite{howellLifeAffirmingBiosensingPublic2019,freyBreezeSharingBiofeedback2018}).
Audio feedback was not considered in our context
because it may cause potential distractions if provided in parallel~\cite{freyBreezeSharingBiofeedback2018} with videos, especially those that possess especially rich auditory expressions~\cite{bannisterDistinctVarietiesAesthetic2019}.

Since we intended to show viewers' affects only at the fringe of attention while they enjoy online videos,
we first envisioned how to transform these collected design examples to convey aggregated frisson data,
and then settled the following design requirements:
\begin{itemize}
  \item[\bfseries DR1] The feedback should reflect either the percentage of viewers experiencing frisson as a single time series,
  and/or every person's own frisson status as separate time series.\label{dr1}
  \item[\bfseries DR2] At any moment, the feedback should reflect at least the magnitude(s) of the frisson time series data in correspondence with the current video progress.\label{dr2}
  \item[\bfseries DR3] The feedback should cause minimal distraction to users' main focus on video-watching.\label{dr3}
  \item[\bfseries DR4] The feedback does not require viewers to take any actions.\label{dr4}
  \item[\bfseries DR5] The feedback's relevance to frissons should be easily understandable.\label{dr5}
\end{itemize}

Furthermore,
we adapted the framework of peripheral design by Pousman and Stasko~\cite{pousmanTaxonomyAmbientInformation2006} to help distill the essence of the design examples
and guide the organization of our own designs in the following steps.
The framework has four dimensions:
\emph{informational capacity} (i.e., general trends of viewers' physiological states),
\emph{notification level} (i.e., extent of awareness without interrupting the main task),
\emph{representational fidelity} (i.e., precision of patterns encoding the data from the world),
and \emph{aesthetic emphasis} (i.e., perceived importance given to aesthetics)~\cite{pousmanTaxonomyAmbientInformation2006}.
We required viable design candidates to have a proper degree of information capacity and notification levels as well as a basic aesthetic emphasis.
In other words, we only kept the design alternatives that have the potential to show and only show the thrills of other viewers in a sensory-comfortable manner without distracting video-watching.
We also modified the representational fidelity scales of the framework as follows,
because the original framework only covered visual designs and did not specialize in
the representations of affective information:
\begin{enumerate}
  \item[1.] \textbf{Haptics}: haptic patterns that reflect the frisson moments,
  \item[2.]  \textbf{Abstract patterns}: visual patterns that do not have any semantic meanings but are valid for encoding information related to frisson moments.
  \item[3.]  \textbf{Language symbols}: visual representations of linguistic elements (e.g., letters, numbers, or punctuation marks) commonly associated with the frisson sensation.
  \item[4.]  \textbf{Metaphorical icons}: visual metaphors that use analogy or association to illustrate frisson experiences.
  \item[5.]  \textbf{Pictorial Icons}: visual icons that reassemble the physical reactions or relevant physical attributes at frisson episodes.
  \item[6.]  \textbf{Indexical visualizations}: visualizations (e.g., plots) of raw numerical values of frisson-related biosignals.
\end{enumerate}

A separate category of haptics was added as we found it hard to place haptic feedback on the representational fidelity scale~\cite{pousmanTaxonomyAmbientInformation2006}. Take vibration as an example. Some individuals may consider it as an abstract notification without meaning, while others feel that it directly mimics the tingling feeling of the skin.

\begin{table*}[h]
  \caption{The design candidates after the brainstorming session and the follow-up organization}
  \label{tab:brainstorm-design}
  \begin{tabularx}{\textwidth}{@{}>{\bfseries}p{2cm}Xp{2.4cm}p{1.4cm}@{}}
    \toprule
    Names of the Designs\textsuperscript{\S} &
      Feature Description &
      Representational Fidelity &
      Aesthetic Emphasis\textsuperscript{\#}
\\ \midrule
    Coldness &
      A TEC attached to the skin generates coldness at frisson moments to simulate the cold feeling that often comes with frisson~\cite{heFrissonWavesExploring2022,haarAugmentingAestheticChills2020}. The aggregated frisson data is inversely mapped to the temperature. &
      haptics &
      low \\
    Vibration &
      A motor attached to the skin vibrates at frisson moments. The aggregated frisson data is mapped to the vibration strength. &
      haptics &
      low \\
    Dots &
      Several dots scatter in the background representing the viewers; the dots turn to warmer colors and gather to create an explosion effect at frisson moments. The aggregated frisson data is mapped to the color and the explosion extent.\newline
      \includegraphics[width=1cm]{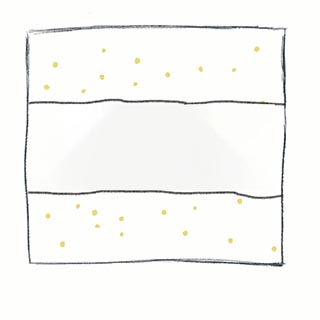}
      \includegraphics[width=1cm]{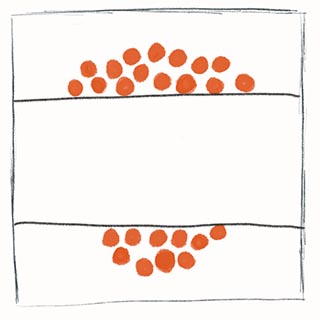}
      \includegraphics[width=1cm]{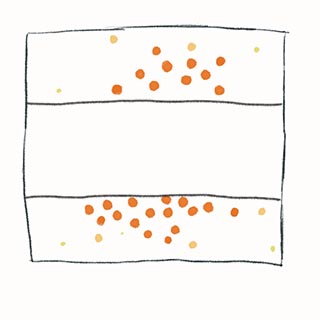} &
      abstract patterns &
      high \\
    Exclamation Mark &
      Inside a region at the bottom right of the video area, particles in the shape of exclamation marks appear and float upward at frisson moments. The aggregated frisson data is mapped to the density of particle emission.\newline
      \includegraphics[width=1cm]{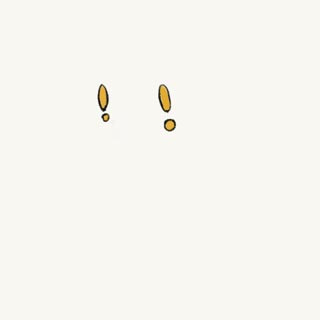}
      \includegraphics[width=1cm]{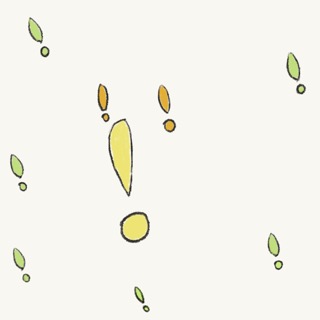} &
      language symbols &
      mid \\
    Texture &
      Dense textile texture gradually appears (and then disappears) on the background at frisson moments to represent the scratchy feeling of the skin. The aggregated frisson data is mapped to the opacity of the texture pattern.\newline
      \includegraphics[width=1cm]{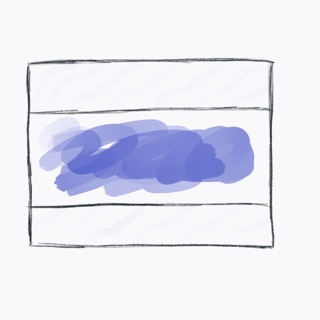}
      \includegraphics[width=1cm]{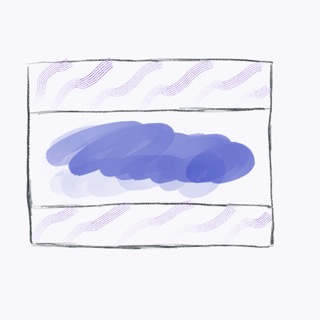} &
      metaphor &
      high \\
    Thermometer &
      A thermometer icon resides at the bottom right corner of the video area. The aggregated frisson data is mapped to the height of the liquid inside.\newline
      \includegraphics[width=1cm]{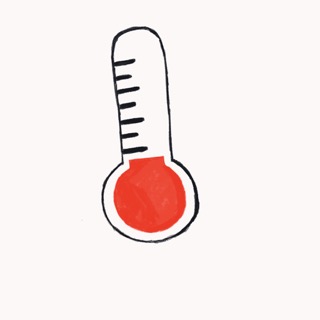}
      \includegraphics[width=1cm]{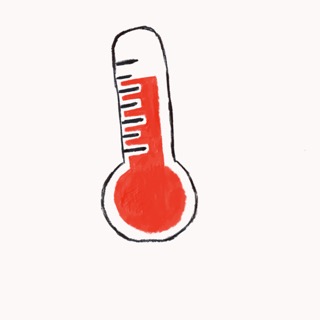} &
      pictorial icons &
      mid \\
    Visualizer &
      Several height-changing bars are placed at the bottom of the video area like music visualizer software. The individuals' frisson data is mapped to the height of each bar.\newline
      \includegraphics[width=1cm]{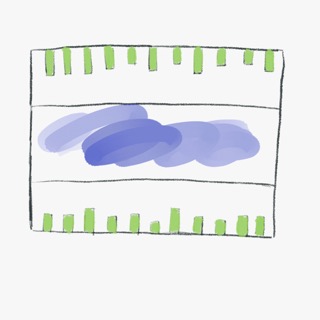}
      \includegraphics[width=1cm]{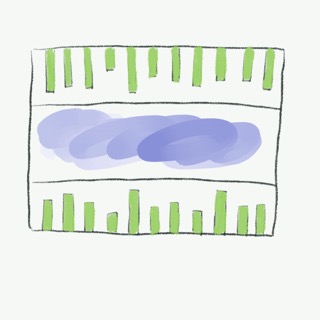} &
      raw visualization &
      high \\
    Line Plot &
      A line plot reflects the aggregated frisson data. It moves inside a clip region at the bottom right corner of the video area, so that only the portion around the current time window is visible.\newline
      \includegraphics[width=1cm]{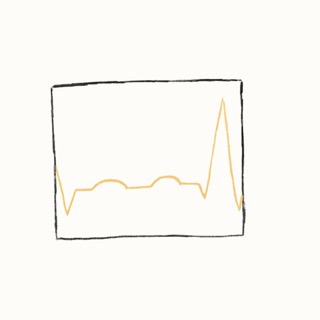}
      \includegraphics[width=1cm]{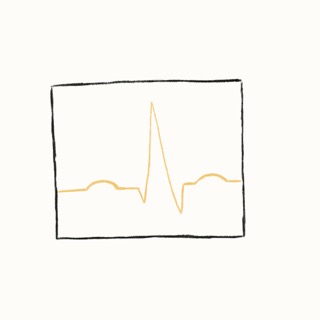} &
      raw visualization &
      low \\
    \midrule
    \multicolumn{4}{l}{\textsuperscript{\S}: We do not provide sketches of non-visual designs. We only describe them verbally.} \\
    \midrule
    \multicolumn{4}{>{\raggedright\arraybackslash}p{\linewidth}}{\textsuperscript{\#}: Originally, this is an unlabelled and continuous axis
    on a whiteboard canvas where researchers put design candidates.
    WLOG we assign three labels, ``low'', ``mid'', and ``high'',
    for clearer presentation.} \\
    \bottomrule
  \end{tabularx}
\end{table*}

Then, we used these categorized design examples from literature as seeds and conducted a brainstorming session to propose varieties of frisson feedback designs for simulating group viewing of online videos.
We first independently put down possible design alternatives. Apart from conforming to the design requirements,
we ensured to generate enough candidates to cover
every representational category listed above.
Afterward, we modified, merged, or removed the initial designs together through discussions.
Such a process resulted in the design candidates shown in Table~\ref{tab:brainstorm-design}.
For each listed design, we also reached a consensus on its \emph{aesthetic emphasis} (low, medium, high) in the discussion.

\subsubsection{Focus Group Interview}\label{sec:design:process:interview}
To evaluate the design alternatives yielded in the brainstorming session from a user's perspective, we conducted a focus group interview with seven participants (4 male and 3 female with an age range of 23--27) from a local university through word-of-mouth.
These participants were frequent users of online video platforms (i.e., at least once a week) and had adequate HCI experience.
During the interview, we asked them to name the genres of videos that often give them frisson moments at which they would like to resonate with others.
Next, we presented animated sketches of each design candidate (shown in Table~\ref{tab:brainstorm-design}) and invited the focus group to picture their effects during video-watching. We encouraged the participants to discuss the suitability of these possible designs in reflecting other viewers' frisson moments collectively by outlining their advantages and shortcomings~(see the supplementary materials).
The focus group could also propose amendments or derivations
from existing designs with justifications.
Lastly,
we asked each participant to eliminate unsatisfactory designs by majority voting and arrived at the three final designs:

\begin{enumerate}
  \item[1.]  \textbf{Ambient light}: Derived from the \textbf{Dots} design and inspired by stage lights.
  An ambient light effect is attached to the boundary of the video and glows at viewers' frisson moments.
  The group frisson data is mapped to the magnitude of the ambient light
  \item[2.] \textbf{Vibration}: Explained in Table~\ref{tab:brainstorm-design}. Due to participants' disinterest in too salient vibration feedback
  and potential confusion with push notifications from smart watches,
  we attach the vibration motor to users' fingertips.
  \item[3.] \textbf{Icon}: Similar to the \textbf{Thermometer} design explained in Table~\ref{tab:brainstorm-design},
  where an icon animates at peripheral vision to represent the frisson data.
  However, due to its relation with temperature,
  some participants commented a thermometer might only be suitable for
  videos with passion and ardor
  or when it reflects body temperature data.
  Thus, we conducted a follow-up iteration of the icon design (see Section~\ref{sec:design:icon}).
\end{enumerate}

In summary, feedback from the focus group helps us further refine the proposed design requirements.
In particular, for~{\bfseries DR3} 
\textbf{(minimal distraction)}, a proper design should not have too many animating parts. It cannot be too subtle either; one should be able to notice the animation without directly looking at it.
As for~{\bfseries DR5} 
\textbf{(good understandablity)}, a design should not merely denote a specific emotion category because frisson may be caused by various feelings like awe, surprise, pleasure, etc.~\cite{bannisterDistinctVarietiesAesthetic2019}.
Also, a design for instant affective communication should avoid using analytical plots.
This aligns with previous findings in~\cite{liuCanBiosignalsBe2017}.

\subsubsection{Iterative Icon Design}\label{sec:design:icon}

Regarding the interviewees' concerns about the thermometer icon design,
we retained the icon feedback but explored other choices of icons in the following iteration.
We started by summarizing several design requirements specifically
for this icon design (DR\textsubscript{I} for short) from
the interviewees' feedback:

\begin{enumerate}
  \item[DR\textsubscript{I}1] The icon should be clearly visible on top of the video.\label{dri1}
  \item[DR\textsubscript{I}2] Its meaning can be efficiently understood as referring to frissons.\label{dri2}
  \item[DR\textsubscript{I}3] The animating part of the icon is salient enough.\label{dri3}
  \item[DR\textsubscript{I}4] The transition should be salient and rapid.\ (cf.\ the \textbf{Exclamation Mark} design, changes in the density of the emitted particles are hard to identify instantly until previous on-screen particles float away and get destroyed.)\label{dri4}
  \item[DR\textsubscript{I}5] The icon is aesthetically pleasing.\label{dri5}
\end{enumerate}

Then,
we explored alternative icons and their animations to
meet the above requirements
and semantically relate to frisson.
Since frisson itself is an abstract psycho-physiological reaction,
we paid particular attention to the physical and psychological reactions
that often occur with frisson and are easier to depict through icons,
like shivering~\cite{craigExploratoryStudyPhysiological2005},
hair standing~\cite{kimBreathingFrameInflatableFrame2015},
skin tingling~\cite{craigExploratoryStudyPhysiological2005},
and being aroused~\cite{goldsteinThrillsResponseMusic1980}.
The complete list of our alternatives is included in the supplementary materials.
Eventually,
we chose the lightning bolt icon,
for it can indicate the skin's tingling feeling
as if electric currents are running through.
The size-changing animation is also salient enough.

After we settled on the icon design,
the final three frisson feedback designs (\textbf{Ambient Light}, \textbf{Icon}, \textbf{Vibration})
are illustrated in Figure~\ref{fig:teaser}.

\subsection{Selecting Online Videos}\label{sec:design:video}

We applied a non-trivial selection process of frisson-triggering videos to ensure that
our sample videos were likely to trigger frisson.
Note that it is out of this paper's main scope to
discover what types of online videos best trigger frisson.
The selection process only ensures that the videos can
draw the majority viewers' attention and interests~\cite{colverGettingAestheticChills2016,silviaPersonalityPiloerectionIndividual2011},
and assist the subsequent evaluation of this frisson-sharing channel.

We first select several video categories according to literature
and the focus group interview (see Section~\ref{sec:design:process}).
Then, we focused on the video types that commonly spanned between three and seven minutes,
which is the most common video length on online platforms~\cite{ahsanCharacterizingInternetVideo2014}.
The result candidate categories are:
music performances,
scenery,
movie mashup,
anime mashup,
and game mashup (both gameplay and narrative footage included).

Then,
two researchers who self-reported to have frequently experienced frisson when watching online videos
independently selected 1--3 videos on YouTube and Bilibili
that fell into each category.
The researchers then cross-viewed the videos not selected by themselves to
pick those perceived as frisson-triggering by both.
The final videos include two music performances
(one orchestra\footnote{\url{https://www.youtube.com/watch?v=w_TaNm6lccQ}} and
one pop music\footnote{\url{https://www.youtube.com/watch?v=zsZMeEl6Gxk}}),
one superheroes/-heroine
movie mashup\footnote{\url{https://www.bilibili.com/video/BV1Vb41177ud/}},
and one anime mashup\footnote{\url{https://www.bilibili.com/video/BV1bK411J7kX/}}.
The orchestra video was played since 1:40 to prevent being significantly longer than other videos.
The validity of our video selection is further confirmed in the following stage.
When we collected viewer's biosignal data when watching the four videos (see Section~\ref{sec:design:data}),
we also asked them to self-report which videos have triggered frisson.
All four videos were reported to have triggered frisson for some viewers,
and corresponding patterns were identified in their biosignal data.

\subsection{Collecting Biosignal Data}\label{sec:design:data}

After we have selected sample online videos,
this subsection describes the following steps of
sensor implementation,
biosignal data collection,
and the data processing pipeline.

\subsubsection{Sensor Implementation}\label{sec:design:data:sensor}

Sudden rises in EDA signal are the most efficient biosignal predictors for frissons~\cite{craigExploratoryStudyPhysiological2005,neidlingerAWElectricThatGave2017}.
We originally intended to use existing commercial smart wearables
since it could better simulate real-world deployment.
However, current commercial smart wearables do not provide
API for high-frequency EDA data acquisition.
Therefore, we eventually developed a custom sensor pipeline
using a Grove GSR sensor\footnote{\url{https://wiki.seeedstudio.com/Grove-GSR_Sensor/}},
a Raspberry Pi 3B+,
and an analog-to-digital converter.
The EDA data was sampled at 5Hz,
broadcast to other parts of the system in real time
with the MQTT protocol,
and saved in a local MongoDB time series collection.
Another browser extension was written to
listen to the play and stop events of the video element on the webpage.
The event type (play or stop) was sent to the same database
together with the timestamp of the event.
The implementation of the features above
is illustrated in Figure~\ref{fig:impl}
in Section~\ref{sec:design:impl},
together with other parts to be introduced in the following sections.

\subsubsection{Data Collection Procedure}\label{sec:design:data:procedure}
\begin{figure}[h]
  \centering
  \begin{subfigure}{0.48\linewidth}
    \includegraphics[width=\linewidth]{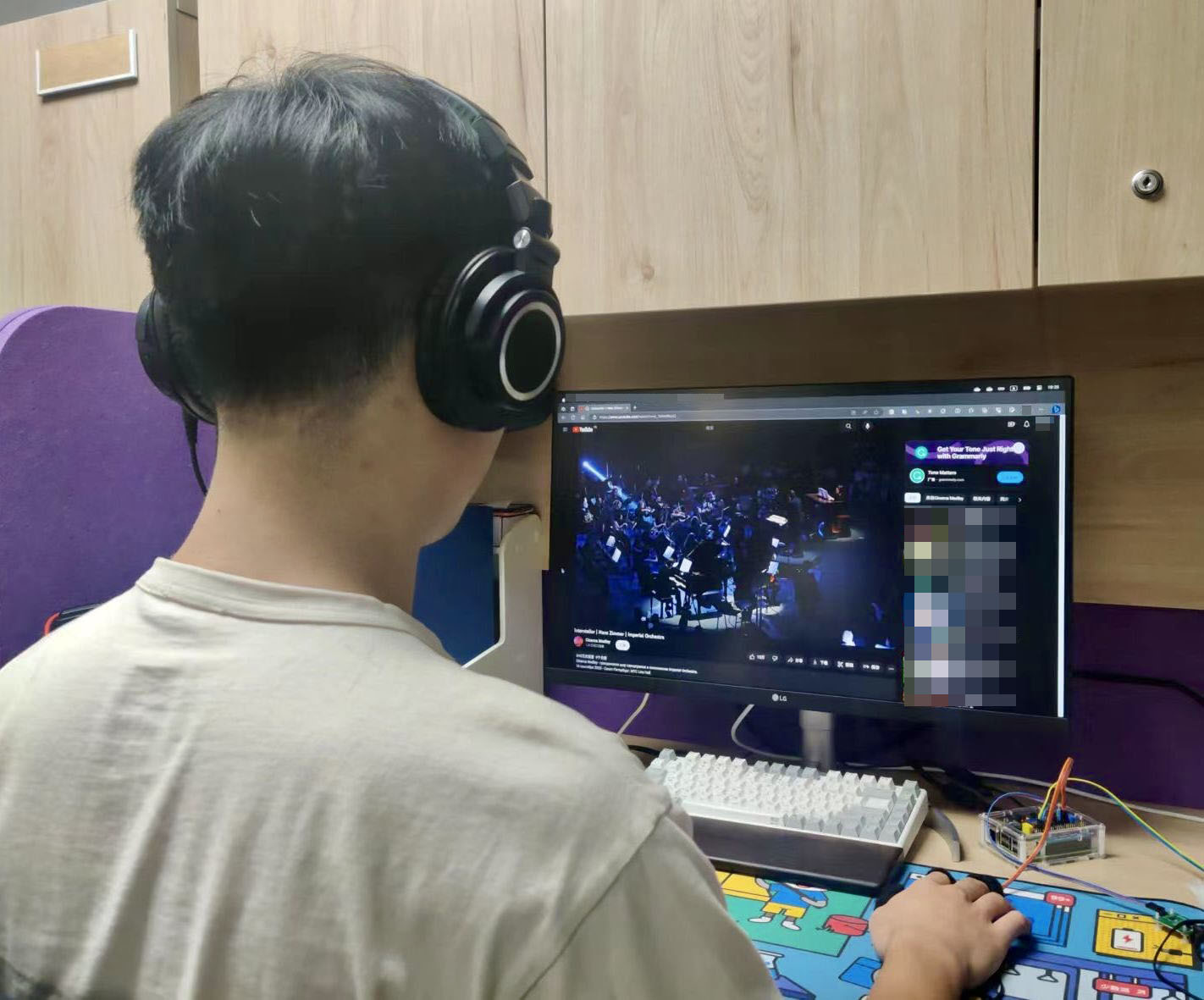}
    \Description{A person wearing headphones and fingertip sensors is watching a YouTube video through a monitor on a clean working desk.}
    \caption{The third-person view.}
  \end{subfigure}
  \hfill
  \begin{subfigure}{0.409\linewidth}
    \includegraphics[width=\linewidth]{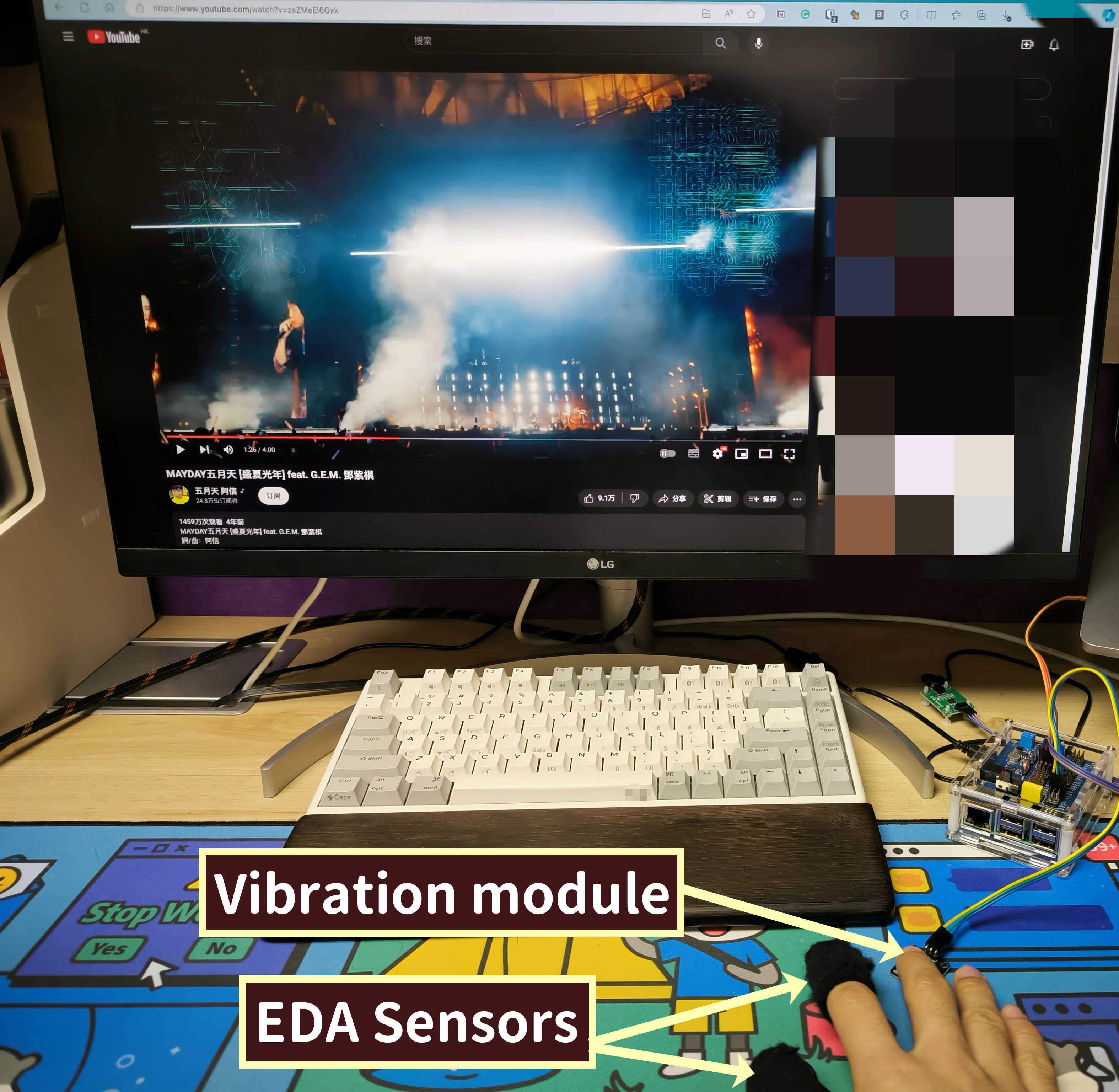}
    \Description{A hand placing on a table, and a monitor is on the table playing a YouTube video. EDA sensors are attached to the thumb and the index finger. The middle finger places on a vibration module.}
    \caption{The first-person view.}
  \end{subfigure}
  \caption{One researcher showcasing the video-watching and sensor setup.}
  \label{fig:data-collection}
\end{figure}

With the IRB's approval,
we recruited 20 people (10 male and 10 female)
from a local university.
Each participant watched all four sample videos
with the EDA sensor
and the same pair of Audio-Technica M50xBT2 headphones (in wired mode to minimize delay).
All extra features on the video platforms were turned off,
and the webpages were in dark and non-fullscreen mode.
\edit{The setup is shown in Figure~\ref{fig:data-collection}.}

The orders of videos was adjusted using
balanced latin square (4 combinations, 5 participants each).
Participants were informed of potential uncomfortable content in the videos in the consent form,
and they were allowed to take short breaks between videos
if they felt tired or hard to keep focused.

\subsubsection{Data Processing}\label{sec:design:data:processing}

To provide an overview of the audience's collective affective states
and their synchrony of frisson experience,
we aggregated the EDA signals of the 20 individuals by calculating the mean value of their frisson series.
The aggregation could also relieve the privacy concern of the identifiability of biosignals~\cite{robinsonChatHasNo2022}.
But prior to the aggregation,
the raw sensor data was processed in four steps to derive their frisson series.
Firstly,
we used a low-pass filter with a sample rate of 30 Hz to eliminate the sensor's own noise and improve the SNR (signal-to-noise ratio).
Then,
to eliminate the baseline drift of biosignals caused by natural changes in skin hydration and temperature,
we estimated the baseline using a moving average filter with a window size of 50, and subtracted the baseline from the original signal~\cite{chouhanTotalRemovalBaseline2007}.
A normalization followed to scale the data into a common range of 0--1.
This step ensured all data points contributed equally to the aggregation and analysis.
Finally, we derived everyone's frisson data via peak detection and signal quantization.
A frisson is of short duration and triggers an increase in the EDA signal~\cite{craigExploratoryStudyPhysiological2005}.
Therefore, we used a peak detection algorithm with time window of 5 seconds and minimal prominence of 0.6.
Based on the peak detection results, individual analog EDA signal is quantized into a binary frisson series.

\subsection{Implementing Frisson Feedback}\label{sec:design:impl}

\begin{figure}[h]
  \centering
  \includegraphics[width=\linewidth]{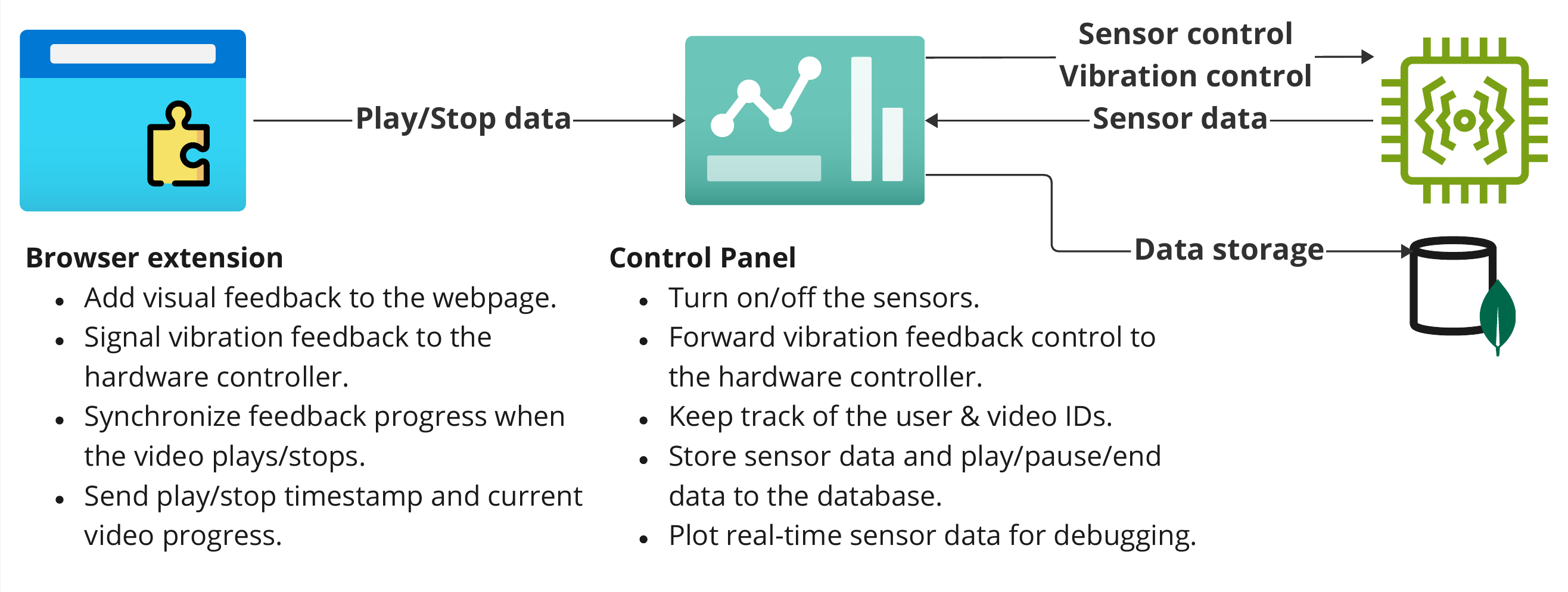}
  \caption{A diagram of the implementation,
  \edit{elaborated in Sections~\ref{sec:design:data:sensor} and \ref{sec:design:impl}.}}\label{fig:impl}
  \Description{Four parts are interconnected in the diagram: Browser Extension, Control Panel, Database, and Hardware. Browser Extension sends play/stop data to the Control Panel, and the Hardware sends sensor data to the Control Panel. The Control Panel sends all the data it receives to the Database. The Control Panel controls the hardware, including the sensors and the vibration motor.}
\end{figure}

To better simulate real-world user experience,
we implemented the \textbf{feedback module} on existing online video platforms (YouTube and Bilibili).
We extended the browser extension mentioned before in the \textbf{sensor module} (see Section~\ref{sec:design:data:sensor})
to modify the video page's content.
Figure~\ref{fig:impl} shows a diagram of the complete implementation.

The browser extension
can load the aggregated frisson time series data.
It contains a popup window,
where researchers can turn on or off a frisson feedback
on the current video page.
For the \textbf{Ambient Light} and the \textbf{Icon} designs,
the extension locates the video element on the current webpage.
The \textbf{Icon} feedback will create an icon element and
position it accordingly.
We use the \verb|anime.js| library to initialize
an animation timeline from the aggregated frisson time series data.
The animation progress is synchronized with the video playback progress.
For the \textbf{Vibration} feedback,
a PWM-adjustable vibration motor was connected to the Raspberry Pi,
and the browser extension sent the vibration strength
corresponding to the current video playback progress
to the Raspberry Pi with the MQTT protocol.
The vibration motor we used had a maximal
speed of 6000RPM\@.
To avoid excessive distraction,
we lowered the vibration strength
and linearly mapped the frisson data to
only 0\%--70\% of the maximal vibration strength.

\section{Evaluation}\label{sec:eval}

This section describes the evaluation process of the
frisson-sharing designs and the results.
We conducted a mix-methods, within-subject study
that mainly evaluated three aspects:
1) the efficiency and outcomes of affective communication through the frisson feedback,
2) the social relationship between the participant and other viewers,
3) the usability of the feedback.
The first two aspects respond to {\bfseries RQ2} 
 and the third aspects respond to {\bfseries RQ1}
. {\bfseries RQ3} 
 will be respond to in Section~\ref{sec:dc},
where we summarize design considerations from the evaluation results.

\subsection{Protocols}

\subsubsection{Participants}
With the IRB's approval,
we recruited 48 people (26 male, 21 female, and 1 non-binary
with an age range of 18--30 and ID P1--P48)
from a local university through word-of-mouth and social media advertisements.
The user study adopted the same monitor and audio device setup
as the data collection procedure (see Section~\ref{sec:design:data:procedure}),
and the video platform's webpage configurations are also the same.

\subsubsection{Procedures}
Each participant watched all four videos and experienced all three feedback designs
plus one baseline condition with frisson feedback turned off.
Both the order of the feedback features (three designs and one baseline)
and the order of the four videos was adjusted using balanced Latin square,
respectively.
The resulting four feature orders and four video orders
formed 16 combinations,
and three participants were assigned to each combination.
After watching each video,
participants completed a questionnaire;
after all four videos,
they were invited to a semi-structured interview.
The entire procedure took around one hour,
and each participant received approximately 8 USD equivalent of local currency
as stipends.

The questionnaire followed the three aspects mentioned above.
To evaluate how efficiently the frisson information was conveyed
through such designs,
we adapted the ``Interpersonal --- Comprehension'' part
of the Short Profile of Emotional Competence (S-PEC) questionnaire~\cite{mikolajczakMeasuringIntrapersonalInterpersonal2014}.
We added custom questions about some potential outcomes of receiving others' frisson information,
as were collected from the focus group interview (see Section~\ref{sec:design:process:interview}).
Regarding the social relationship between the participant and other viewers,
we evaluated other viewers' social presence using the Inclusion of Other in the Self (IOS) Scale~\cite{aronInclusionOtherSelf1992},
and whether they felt as if watching videos with others together.
People's attribution of their affective states
is also another cue for social information~\cite{vandoornDerivingMeaningOthers2015}.
Hence,
we also asked whether they attribute the presence or the absence of frissons
to the video content or other viewers.
Finally,
the usability was evaluated with two questionnaires.
The scales by Mankoff et al.\@~\cite{mankoffHeuristicEvaluationAmbient2003}
evaluated three feedback approaches
as peripheral designs ---
whether they efficiently delivered information with minimal attention.
The general usability was tested using
the Short Version of the User Experience Questionnaire (UEQ-S)~\cite{schreppDesignEvaluationShort2017}.

During the interview,
we invited participants to elaborate on their comments on different designs,
the overall experience of watching videos with such real-time frisson feedback,
and their attitudes towards this means of communication on online video platforms.

\subsubsection{Data Analysis Methods}
Except that the IOS Scale is in 7 points~\cite{aronInclusionOtherSelf1992},
all the other standardized scales and our additional questions mentioned below are in 5 points~\cite{mikolajczakMeasuringIntrapersonalInterpersonal2014,mankoffHeuristicEvaluationAmbient2003,schreppDesignEvaluationShort2017}.
The affective communication and usability sections
(Section~\ref{sec:comm} and \ref{sec:use})
were omitted for the baseline condition,
and their responses were compared pairwise among the three feedback designs.
The responses to the social relationship section (Section~\ref{sec:social}) were compared between each feedback design and the baseline condition.
These comparisons were conducted using the Kruskal-Wallis test followed by the Dunn's test as the post-hoc analysis.
For our analysis methods,
the a priori power analysis suggested
sample sizes of 42 and 48
(number of groups = 3 or 4,
\(\alpha=0.05\),
\(1-\beta=0.8\),
effect size = 0.5),
which determined our final sample size of 48.
A post-hoc power analysis was also conducted for verification purposes.
If the null hypothesis of the Kruskal-Wallis test was accepted (i.e., no significant difference was found from the comparison),
Dunn's test would be skipped, and no statistics would be reported.

\subsection{Affective Communication and Affective Contagion}\label{sec:comm}

\begin{figure*}[h]
  \centering
  \includegraphics[width=\textwidth]{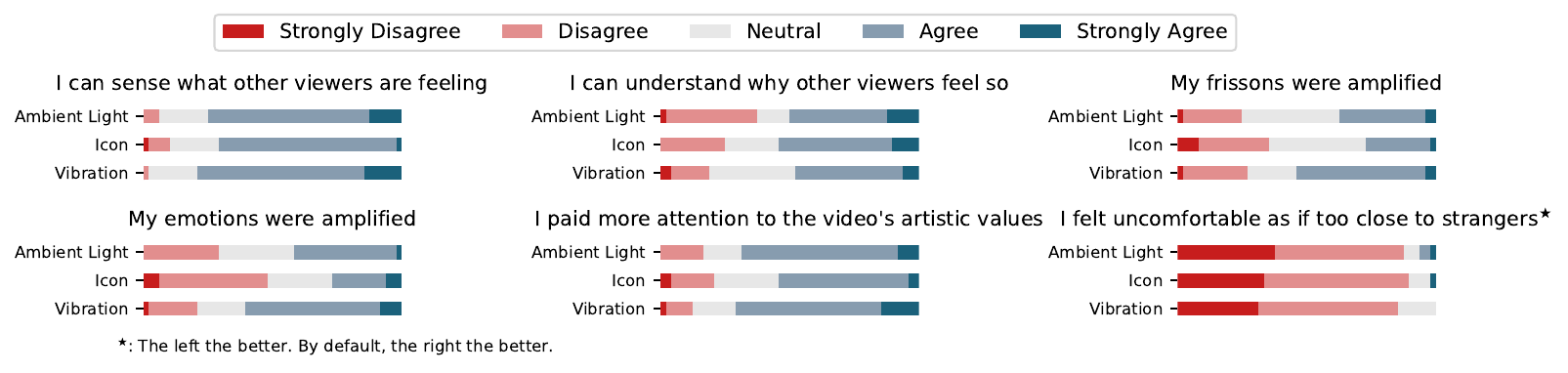}
  \caption{Responses to the ``affective communication'' section of the questionnaire.}
  \Description{In the bar plots, the majority agreed that they could sense what other viewers are feeling, understand why other viewers feel so, and the feedback made them pay more attention to the video's artistic values. Similar numbers of participants agreed and disagreed that their frissons and emotions are amplified. Only four felt uncomfortable as if too close to strangers.}
  \label{fig:eval:comm}
\end{figure*}

In this section,
we compared the questionnaire responses of the three feedback designs.
As is shown in Table~\ref{tab:eval:comm},
the result did not exhibit much statistically significant difference between the three designs.
Figure~\ref{fig:eval:comm} shows the detailed ratings.

\begin{table*}[h]
  \caption{The statistics of the ``affective communication'' section of the questionnaire.
  (Only the questions with null hypothesis rejected are presented.
  The Sig.\ values have been adjusted by the Bonferroni correction.
  \textbf{bold}: Sig. $\leq 0.05$.)}
  \label{tab:eval:comm}
  \footnotesize
  \begin{tabular}{>{\raggedright\arraybackslash}p{3cm}cccccc}
  \toprule
    \multirow{2}{*}{Quetions (shortened)} & \multicolumn{2}{c}{Ambient Light VS Icon} & \multicolumn{2}{c}{Icon VS Vibration} & \multicolumn{2}{c}{Ambient Light VS Vibration} \\ \cmidrule{2-7}
                      & Std.Stat.   & Sig.   & Std.Stat.   & Sig.  & Std.Stat.   & Sig.\\
                      & Effect Size & Power  & Effect Size & Power & Effect Size & Power\\ \midrule
  My emotions were amplified & 1.73 & 0.254 & -3.33 & \textbf{0.003} & -1.61 & 0.323 \\
                              & --- & --- & 0.623 & 0.856 & --- & --- \\ 
  \bottomrule
  \end{tabular}
\end{table*}

\subsubsection{Affective Communication Efficiency}\label{sec:comm:use}

Overall,
sensor-based frisson sharing afforded more efficient affective communication.
From the senders' perspective,
several participants were open to sharing their feelings
with other video viewers,
but were hindered by the inefficiency of existing text-based channels.
The sensor-based approach could minimize the cognitive load and interruption to
organize languages (P10) and
type (P7, P9, P10, P27, P34, P41, P44, P46).
On the receivers' side,
the feedback of the audience's aggregated data
was straightforward (P37, P40, P47) and lightweight (P39).
P40 commented,
\textit{``To me, `stronger affective response' is one category of similar information.
I think it is good to group and abstract these messages like this
and show an overall percentage.''}
Some participants are especially interested in other viewers' affective sharing,
and such a feature helped filter other information (P36, P41).
The absence of text could also prevent uncomfortable, hostile arguments (P48).
The direct reflection of others' physiological responses
also appeared more authentic (P47).
However,
contrary to the conciseness of non-text feedback,
the absence of text is also a shortcoming to some other participants.
It could only communicate frissons instead of richer emotions (P31).
And since it aggregated everyone's frisson experience,
one might feel hesitant about the significance of their own marginal contribution (P26, P47, P48).
P47 explained,
\textit{``I have less frequent physiological responses to online videos.
That's why I'm not very into this form of communication.''}

We intended to use biofeedback like frissons as a proxy of viewers'
emotional arousal.
However, the disparity between physiological and psychological responses
resulted in the wrong interpretation of the information by some participants.
For example,
the presentation of the percentage of viewers having frissons
resulted in non-binary data,
while individual frisson experiences are more like instant and binary data.
Therefore,
some participants mistook it for emotional arousal,
which itself was non-binary (P26, P41).
P41 added,
\textit{``Although frissons are instant,
the accompanying emotions are not.
When I am still in the mood,
it is annoying to see the data quickly drop.''}
Some participants also tended to mistake the sharing of natural frissons
for a stimulus for synthetic frissons (P7, P31).
Especially for the vibration,
some praised it for its straightforward resemblance to frissons (P42, P43),
while some regarded it as a stimulus to trigger frissons (P15, P32, P40).
P40 said,
\textit{``The vibration was like the UX augmentation installed on cinema seats,
so I expected it to vibrate at every exciting moment, but it was not the case.\ldots{}
When I felt excited but it didn't vibrate, I would become confused.''}

\subsubsection{The Impact on Affective Experience}\label{sec:comm:affect}

The sense of social bonding with other audience members
can influence the affective contagion and, ultimately,
their own affective responses~\cite{garridoAudienceEngagementCommunity2020}.
In Table~\ref{tab:eval:social},
participants also additionally attributed their frisson experience
to the presence of other viewers' frisson information.
It was only a secondary factor, though,
as all participants still reported that
video content was the main cause of their frissons.
Similar to the sense of togetherness,
the changes in the attribution ratings were less significant
in the two concert videos.

Besides the direct transmission of physiological reactions,
the impact of perceiving others' frissons was also emotional.
Especially when the participant's frisson moments coincided with others,
the accompanying emotions are often amplified
(P1, P3, P15, P26, P27, P29, P37, P39, P40, P41, P42, P44, P47).
However, a downside was additional negative emotional experiences
when participants noticed mismatches between their frissons and others,
adversely affecting their video-watching experiences.
The negative feelings included
interruption in their emotional experiences (P19, P41),
regret at the lack of resonance (P24, P29, P41, P44),
and self-questioning (P17).
P29 commented,
\textit{``Such negative feelings are stronger than text-based communication.
Because I know that this feedback includes the entire audience's reaction.
[If it shows nothing,] then it means absolutely no one feels the same as I do.''}

Seeing the nuanced attitudes and interpretations of the audience's
frisson data,
we asked participants whether they saw
this way of sharing their spontaneous physiological reactions
as self-expression of personal interests, aesthetic preferences, etc.
Most participants did not,
as they could not see profound and expressive elaborations from mere frisson data (P6, P28, P42, P43),
plus individual self-expression dissolved in the aggregated data (P26, P32).
However, around one-third of participants did have such an interpretation
(P1, P17, P19, P21, P27, P29, P30, P31, P33, P34, P37, P40, P45, P47, P48),
as some emphasized their tendency towards frissons
when they were interested in the video (P37, P48),
and some were more inclined to share their feelings about a video
that is arousing (P30, P47).
But for those reluctant to share their frissons,
such an interpretation, if present, was also one reason for
their fear of being judged (P29),
or indifference to strangers' affective information (P31, P45).

\subsubsection{Non-Affective Consequences of Affective Communication}

The sharing of affective information did not always lead to
affective responses.
Non-affective consequences of frisson sharing were also identified among the participants.
48\% of participants stated in the questionnaire that perceiving others'
frissons encouraged them to pay more attention to understanding the artistic values of the video.
P5 added,
\textit{``To some extent, I will try to understand [other viewers]
and put myself in their shoes [when I noticed their frissons].
If the video is something trendy recently,
I don't want to miss out.''}

In addition,
some participants were more interested in observing others'
reactions on online video platforms instead of
resonating with others.
For them, the frisson feedback would be more valuable
if it could accurately reflect the percentage of
viewers having frissons
rather than providing a vivid but rough representation (P32, P40).
P32 said,
\textit{``Maybe you can add some quantitative data like percentages\ldots{}
I don't want it to affect me constantly.
I can glimpse it when I need it, and that's enough.
I am interested in the data itself,
not others' sensations.''}
Although the original intention of our design was
to create a stronger collective and affective video-watching experience,
such findings may be insightful for feature designs.

\subsection{The Social Connection with Other Viewers}\label{sec:social}

\begin{figure*}[h]
  \centering
  \includegraphics[width=\textwidth]{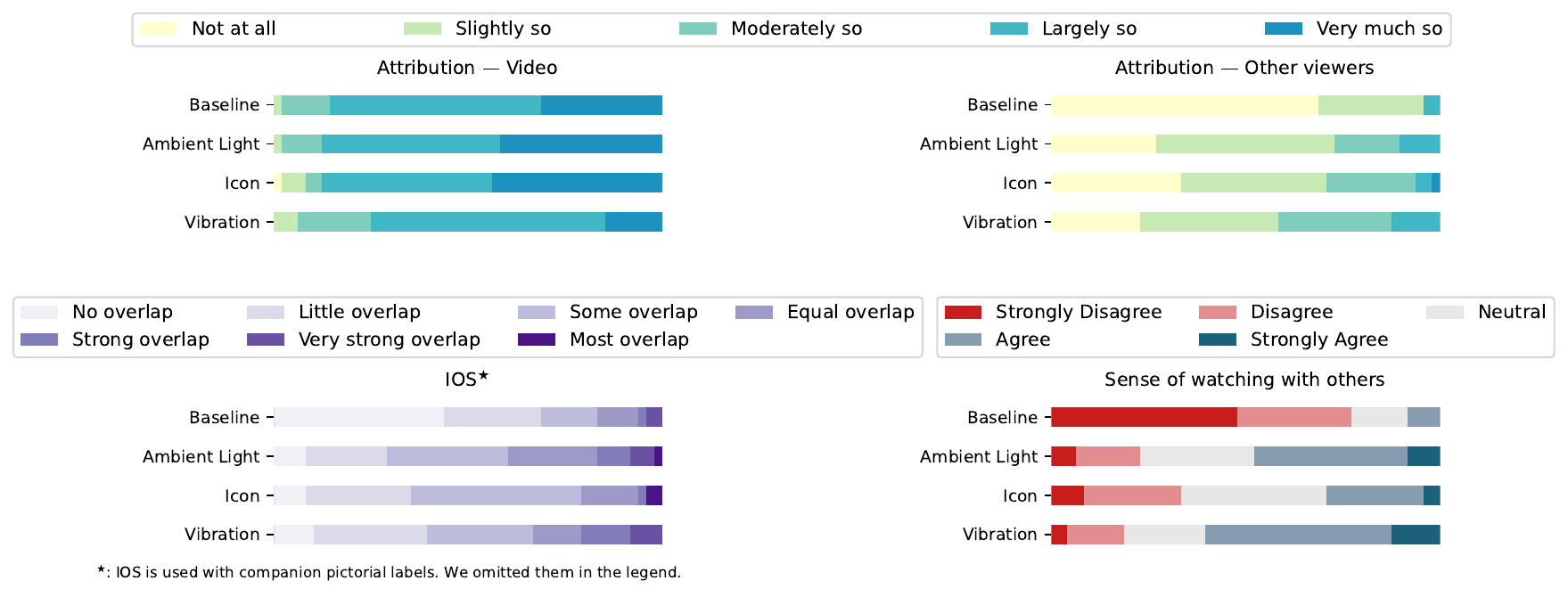}
  \caption{Responses to the ``social'' section of the questionnaire.}
  \Description{In the bar plots, the majority attribute their frisson experience to the video content no matter what. But with frisson feedback is on, an increasing number of participants slightly or moderately attribute to other viewers' reactions as well. An increase in the IOS occurs to a similar extent. The sense of watching with others undergoes a more significant increase compared to the baseline condition.}
  \label{fig:eval:social}
\end{figure*}

In this section,
we compared the questionnaire responses of the baseline condition against each feedback design.
The statistical results are shown in Table~\ref{tab:eval:social} and Figure~\ref{fig:eval:social}.

\begin{table*}[h]
    \centering
    \caption{The statistics of the ``social'' section of the questionnaire.
    Each design is compared to the baseline condition.
    (The Sig.\ values have been adjusted by the Bonferroni correction.
    \textbf{bold}: Sig. $\leq 0.05$.)}
    \label{tab:eval:social}
    \footnotesize
    \begin{tabular}{lcccccc} 
    \toprule
    \multicolumn{1}{c}{\multirow{2}{*}{}} & \multicolumn{2}{c}{Ambient Light} & \multicolumn{2}{c}{Icon} & \multicolumn{2}{c}{Vibration} \\ 
    \cmidrule{2-7}
    \multicolumn{1}{c}{} & Std.Stat. & Sig. & Std.Stat. & Sig.& Std.Stat. & Sig. \\ 
     & Effect Size & Power  & Effect Size & Power & Effect Size & Power\\
    \midrule
    Attribution --- Video & -0.93 & 1.000 & -0.99 & 1.000 & 1.96 & 0.149\\
     & 0.131 & 0.097 & 0.091 & 0.073 & 0.410 & 0.511 \\
    \midrule
    Attribution --- Other viewers & -4.03 &\textbf{0.000} & -3.63 & \textbf{0.001} & -5.05& \textbf{0.000}\\
    & 0.789 & 0.969 & 0.759 & 0.957 & 1.031 & 0.999 \\
    \midrule
    IOS & -4.33 & \textbf{0.000} & -3.10 & \textbf{0.006} & -3.59 & \textbf{0.001} \\
    & 0.832 & 0.981 & 0.638 & 0.871 & 0.766 & 0.960 \\
    \midrule
    Sense of watching with others & -5.79 & \textbf{0.000} & -4.25 &\textbf{0.000} & -6.77 &\textbf{0.000} \\
    & 1.513 & 1.000 & 1.129 & 1.000 & 1.728 & 1.000 \\
    \bottomrule
    \end{tabular}
\end{table*}

\subsubsection{The Sense of ``Watching Together''}

Table~\ref{tab:eval:social} shows that every design managed to
increase the social presence of other viewers
and create a sense of ``watching together''.
Some participants
highlighted that the frisson feedback
was like a company when they watched the videos (P26, P30).

For some participants,
the sense of togetherness was especially triggered
at moments of physiological synchrony or
by specific feedback designs.
For P48,
it was the vibration design:
\textit{``Since the video itself was visual information,
if the additional information arrives in another modality,
it's easier for me to perceive,
and it is more like real people beside me having such reactions to the video.''}
Similarly,
for the visual designs that \textit{``looks more like a part of the video''},
P44 thought,
\textit{``The sense of `being together with other viewers' is weaker.''}
At the moments of physiological synchrony,
participants often felt empathy with other viewers
(P2, P6, P7, P8, P10, P13, P17, P19, P21, P22, P25, P27, P28, P30, P33, P40, P43, P44),
and were happy at such moments (P19, P22, P26, P28, P37, P42, P43, P44).
P47 added,
\textit{``When my goosebump moments concurred with those of others,
it was like another invisible viewer by my side,
and I could feel stronger empathy.''}

In contrast,
the sense of being together with other viewers might sometimes backfire.
When the participant's frisson experiences mismatched
those of other viewers,
it might deliver a feeling of being too close to strangers (P45)
or adversely affect the perceived social relationship with other viewers (P24).
When certain feedback design was annoying,
participants might also blame the ``presence'' of other viewers
for the drop in their user experience (P40).
Some other participants were generally not interested in other viewers' feelings on online video platforms (P20, P31),
or they did not want others' feelings to affect their own (P45).
P20 commented,
\textit{``I don't want to get this information (others' frissons).
That's why I like vibration the most,
because I had the least association between this feedback and others' reactions.''}
P35 also expressed a similar concern,
\textit{``I think the ultimate goal of such additional feedback is to enhance our video-watching experience.
If it deliberately emphasizes too much on the fact of `watching together' and invites us to observe their reactions,
we may lose the original intention of watching videos.''}
These comments reflected participants' different needs for social interactions
on online video platforms.
The perception of sharing frissons as a social interaction
with other viewers was still present,
but they desired more control over the user group and the scenarios
to strike up such interactions.

Apart from the overall social effect,
we also compared participants' feedback when watching each video in Table~\ref{tab:eval:social-video}.

\begin{table*}[h]
    \centering
    \caption{The statistics of the ``social'' section grouped by the videos.
    Each design is compared to the baseline condition.
    (Only the questions with null hypothesis rejected are presented.
    The Sig.\ values have been adjusted by the Bonferroni correction.
    \textbf{bold}: Sig. $\leq 0.05$.)}
    \label{tab:eval:social-video}
    \footnotesize
    \begin{tabular}{lccccccc}
    \toprule
    \multirow{2}{*}{} & \multirow{2}{*}{Video} & \multicolumn{2}{c}{Ambient Light} & \multicolumn{2}{c}{Icon} & \multicolumn{2}{c}{Vibration} \\
    \cmidrule{3-8}
    &  & Std.Stat. & Sig. & Std.Stat. & Sig.& Std.Stat. & Sig. \\
    & & Effect Size & Power  & Effect Size & Power & Effect Size & Power\\
    \midrule
    Attribution --- Other viewers & Anime & -3.22 & \textbf{0.004} & -2.59& \textbf{0.029} & -3.19& \textbf{0.004} \\
    & & 1.330 & 0.875 & 1.126& 0.751 & 1.445& 0.922 \\
    & Orchestra  & -1.93 & 0.162 & -1.50& 0.403 & -3.05& \textbf{0.007}\\
    & & 0.740 & 0.411 & 0.555& 0.091 & 2.045& 0.998 \\
    \midrule
    IOS & Anime & -3.31 & \textbf{0.003} & -2.46& \textbf{0.042} & -2.32& 0.061 \\
    & & 1.438 & 0.920 & 1.109& 0.738 & 1.026& 0.671 \\
    & Superhero & -2.78 & \textbf{0.016} & -2.26& 0.072 & -3.16& \textbf{0.005} \\
    & & 1.274 & 0.847 & 1.208& 0.807 & 1.433& 0.918 \\
    \midrule
    Sense of watching with others & Anime & -4.03 & \textbf{0.000} & -2.75& \textbf{0.018} & -4.33& \textbf{0.000} \\
    & & 2.601 & 1.000 & 1.757& 0.984 & 2.309& 1.000 \\
    & Superhero & -2.12 & 0.102 & -2.50& \textbf{0.037} & -3.10& \textbf{0.006} \\
    & & 0.951 & 0.605 & 1.261& 0.839 & 1.549& 0.952 \\
    & Orchestra & -3.25 & \textbf{0.003} & -1.84& 0.197 & -3.19& \textbf{0.004} \\
    & & 1.574 & 0.957 & 1.049& 0.690 & 2.230& 0.999 \\
    & Pop Music& -2.39 & \textbf{0.050} & -1.35& 0.443 & -3.00& \textbf{0.008} \\
    & & 1.122 & 0.748 & 0.582& 0.276 & 1.521& 0.945 \\
    \bottomrule
    \end{tabular}
\end{table*}

We initially anticipated polarized ratings
for the anime video due to participants' varied
connection to the ACG subculture community.
However,
the results in Table~\ref{tab:eval:social-video} were reversed,
showing the most consistently significant social impact for the anime video
and the least for the two music concert videos.
It was mainly because
the two concert videos had already captured
reactions of the on-site audience.
Some participants could resonate with them
and developed a sense of watching with them
even if no extra information was provided (P38, P48).
P38 said,
\textit{``[The feeling of `watching together' without any feedback] was because the [pop music concert] video was in a stadium with thousands of people.
I saw such a scene and felt as if I was also present at the concert.''}

\subsubsection{Willingness to Self-Disclose}\label{sec:willingness}

Self-disclosure of affects is often considered an intimate action~\cite{bakSelfdisclosureRelationshipStrength2012}.
We also discovered participants' dissimilar attitudes about using such frisson sharing in general,
which indicated whether they intended to establish connections with other viewers.
On the one hand,
around one-third of participants were wishfully willing to share their frissons moments in this way
(P3, P7, P9, P10, P14, P15, P19, P21, P25, P26, P27, P28, P37, P41)
or would like to do so when they were interested in the video (P5, P33, P46),
aroused by the video (P43),
or wanted to watch this video with others (P30).
Some explicitly expressed their wish to deliver their excitement to other viewers to
foster resonance (P7, P43),
or become a part of the community (P37).
On the other hand,
some participants did not want to initiate such sharing at all (P20, P23, P31, P38, P39, P45),
and the other participants did not mind but were not motivated to share their frissons.
They were uninterested in telling the mass public about their affective experiences or reaching resonance
(P20, P23, P31, P34, P35, P38, P39, P40, P44, P45).
Some also felt that affective information is too personal to share with strangers (P34, P39, P44, P46).
P29 even added, \textit{``Sharing my frissons gives me the feeling of
being gazed at or judged by others.''}
These thoughts also led to some participants' idea of disclosing frissons to friends only (P30, P39, P34, P44, P46).
They pointed out that sharing physiological reactions with friends
was a good way to strike extended discussions (P30, P34),
and they could feel stronger resonance and connections with their friends if they received positive responses (P34, P44).

From another perspective,
more participants said they would like to turn on the feedback of others' frisson --- at least using their preferred design.
It echoed most participants' current strategies on online video platforms:
despite the interest in others' reactions and thoughts,
fewer liked to disclose their feelings.
Major reasons were the lack of motivation (P2, P20, P22, P23, P27, P34, P35, P36)
and prevention of disagreement (P35, P40, P48).
In such cases,
They maintained their attitudes towards this frisson-sharing feature.
One exception is P27,
who was reluctant to express herself through traditional text-based features
but was happy with this new approach:
\textit{``I would love to use this thing,
because eventually, I hope I can let others know my feelings when watching videos\ldots{}
This approach seems to not require much effort before others can know my feelings.
It feels different from trying hard to express myself and letting others accept my feelings.
If I try hard to express myself but others don't empathize with me,
I will feel that my efforts are wasted.
But this new approach is effortless.''}

\subsection{Usability and Appropriateness}\label{sec:use}

\begin{figure*}[h]
  \centering
  \includegraphics[width=.9\textwidth]{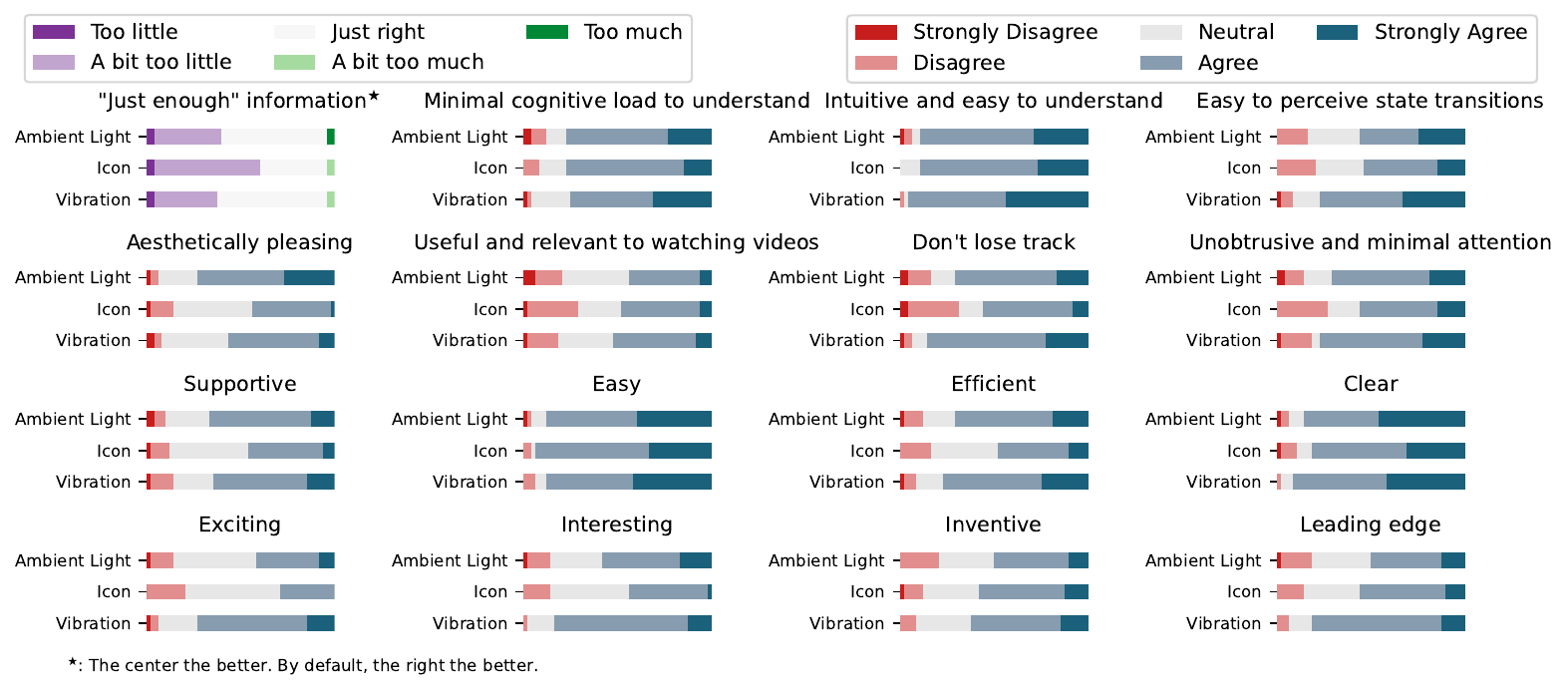}
  \includegraphics[width=0.5\textwidth]{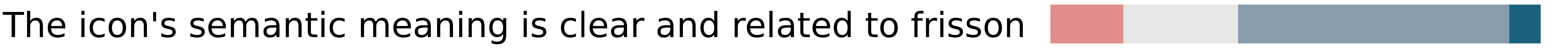}
  \caption{Responses to the ``usability'' section of the questionnaire.}
  \Description{The majority were satisfied with the usability in general, with slightly more neutral ratings of “exciting” and “interesting”. The “‘just enough’ information” question had slightly more than a half rating of “just right”, six too much, and the rest too little.}
  \label{fig:eval:use}
\end{figure*}

In this section,
we compared the questionnaire responses of the three feedback designs using the same approach as Section~\ref{sec:comm}.
The results shown in Table~\ref{tab:eval:use}
did not exhibit much significant difference.
Figure~\ref{fig:eval:use} shows the detailed ratings.

\begin{table*}[h]
    \caption{The statistics of the ``usability'' section of the questionnaire.
    (Only the questions with null hypothesis rejected are presented.
    The Sig.\ values have been adjusted by the Bonferroni correction.
    \textbf{bold}: Sig. $\leq 0.05$.)}
    \label{tab:eval:use}
    \footnotesize
    \begin{tabular}{lcccccc}
    \toprule
    \multicolumn{1}{l}{\multirow{2}{*}{Questions (shortened)}} & \multicolumn{2}{c}{Ambient Light VS Icon} & \multicolumn{2}{c}{Icon VS Vibration} & \multicolumn{2}{c}{Ambient Light VS Vibration} \\\cmidrule{2-7}
    \multicolumn{1}{l}{}                  & Std.Stat.   & Sig.   & Std.Stat.   & Sig.   & Std.Stat.   & Sig.\\
     & Effect Size & Power  & Effect Size & Power & Effect Size & Power\\\midrule
    
    Easy to perceive transitions                       & 0.82        & 1.000      & -2.64       & \textbf{0.025}      & -1.82       & 0.208      \\
     & 0.127 & 0.094  & 0.405 & 0.502 & 0.273 & 0.263\\\midrule
    Aesthetics                       & -1.41       & 0.473      & 3.66        & \textbf{0.001}      & 2.25        & 0.074      \\
    & 0.825 & 0.979  & 0.360 & 0.415 & 0.446 & 0.580\\\midrule
    Do no lose track                            & 1.79        & 0.219      & -3.38       & \textbf{0.002}      & -1.59       & 0.335      \\
    & 0.370 & 0.434  & 0.556 & 0.769 & 0.188 & 0.149\\
    \midrule[\heavyrulewidth]
    Efficient                         & 2.03        & 0.127      & -2.92       & \textbf{0.011}      & -0.89       & 1.000      \\
    & 0.358 & 0.411  & 0.526& 0.723 & 0.157 & 0.119\\\midrule
    Exciting                          & 1.51        & 0.396      & -4.37       & \textbf{0.000}     & -2.87       & \textbf{0.012}      \\
    & 0.355 & 0.406  & 0.959& 0.996 & 0.542 & 0.748\\\midrule
    Interesting                       & 1.88        & 0.182      & -3.92       & \textbf{0.000}      & -2.04       & 0.124      \\
    & 0.308 & 0.321  & 0.871 & 0.988 & 0.403 & 0.498\\\midrule
    Leading                           & -0.45       & 1.000      & -2.58       & \textbf{0.030}      & -2.12       & 0.101      \\
     & 0.061 & 0.060 & 0.470 & 0.625 & 0.517 & 0.708\\\bottomrule
    \end{tabular}

\end{table*}

\subsubsection{Cognitive Load and Attention}

Overall,
participants' preferences varied,
and it was difficult to conclude which feedback design was better.
From Table~\ref{tab:eval:use},
the statistical significance mainly occurred in the comparison
between the icon and vibration design,
where the vibration more saliently showed the changes in
the audience's frisson states,
more accessible to know the audience's current frisson states,
more efficient to use,
and more friendly to use in general.
Many participants elaborated their criticism of the icon design as it
required extra efforts to move their eyes to the corner of the video
(26 out of 48),
and it sometimes mixed up with the video content
(P1, P23--24, P28, P29, P33, P45, P47).
However, there were still participants favoring the icon design,
as it was easy to see (P13, P32)
and obvious when its size changed (P40, P43).

The vibration design was praised by most,
with additional comments on its novel multi-modality user experience (P28, P30, P39).
Nevertheless,
some participants thought it was distracting (P13, P27, P40),
felt forced to perceive the data (P40) or being supervised (P10),
or disliked haptic feedback in general (P31, P37).
As vibration was similar to many existing haptic augmentation of multimedia (see Section~\ref{sec:comm:use}),
the distraction could be especially severe when it mismatched the video's flow (P24, P32).

The ambient light was the second favorite.
It was more visually evident than the icon as it spanned a larger area (P1, P4, P8, P9, P11, P14),
and it was not distracting (P13, P19, P25, P30, P31)
and visually harmonious (P1, P19, P22, P24, P34, P37).
However,
it still required some participants to move their eyes to see (P16, P21, P35, P40, P42, P43, P48),
and some saw it as not evident enough (P28, P33, P43, P45),
or mixed-up with the video content (P21, P28, P33, P44, P45, P47).

In general,
participants did not report satiety from too many stimuli.
However, they still regarded the amount of attention drawn from watching videos
as one critical factor to consider.
Other important factors related to the usability included
the (minimal) efforts to distinguish the frisson data
and the (minimal) extra cognitive load to notice its changes,
despite participants' divergent ratings for each design on these factors.

\subsubsection{Suitability for Different Types of Videos}

In addition to overall cognitive and attentional requirements,
participants also held varying opinions about the
designs' suitability for different types of videos.
They focused on different traits of the sample videos and also preferred different designs to match the traits.

Taking the two mashup videos as an example,
among those who highlighted their strong rhythms,
P33 liked the ambient light,
P21 and P43 liked the icon,
while P35 and P36 liked the vibration.
When underscoring their rich
visual information, P27 favored the two visual designs and P37 the vibration
For the two music performance videos,
some participants preferred the ambient light design
because it resembled the atmosphere of concerts
(P19, P24, P26, P28, P36),
some preferred vibration because it resembled
the impact of powerful music (P27, P44),
or the audience's reactions (P48),
or the awe induced by the music (P43);
some paid less visual attention to musical videos,
and thus liked to see the icon on them (P19, P48).

Instead of exhausting the list,
we identified two key aspects that drove participants' decisions.
One was the video's tone,
which could be magnificent (P33, P35, P44),
exciting (P19, P27, P29, P44),
emotional (P2, P5, 19),
casual (P2, P17, P19, P27),
scary (P44, P48), etc.
The other was the video's content, which could be
plots and twists (P18, P22, P28, P42, P44, P46--47),
art performances (P4--5, P24, P30, P32, P35, P40--41).
competitions (P1, P25, P35, P37),
comedy (P7, P29), etc.
The aspects and traits of interest shall
assist designers in clustering and inferring specific users' preferences.
\section{Discussion}

\subsection{Sharing Frissons as Social Interactions}\label{sec:social-sharing}

In our study,
\edit{we verified that frissons are capable of acting} as an objective and measurable proxy for users' affective states.
\edit{Furthermore, detecting and sharing frissons can be an effective affective communication channel during media consumption such as watching online videos.}
For one thing, people can interpret the affective information embedded in frissons relatively easily.
For another, physiological sensors can capture frisson without additional efforts of manual input.
\edit{Ultimately, presenting the audience's overall frisson data can create a sense of watching together with others
and facilitate asynchronous affective contagion among the viewers.}

From participants' further elaborations of their user experiences,
it is noteworthy that sharing affective information,
including frissons,
is regarded by some participants as a self-disclosure activity,
even though the data they share is combined with others'
into an unidentifiable overview of a larger group.
Such a mindset allows us to explain participants' discussions on their willingness to share their frisson data.
Some participants are especially concerned about the negative impacts of sharing such data,
like stirring up unnecessary disagreement (P40) or
being judged by others (P24, P29)~\cite{andalibiDisclosurePrivacyStigma2020}.
The willingness to self-disclose is also moderated by
the social tie with the other party~\cite{wangModelingSelfDisclosureSocial2016,maAnonymityIntimacySelfDisclosure2016,bakSelfdisclosureRelationshipStrength2012},
the perceived identifiability of the data~\cite{wangModelingSelfDisclosureSocial2016,maAnonymityIntimacySelfDisclosure2016,bakSelfdisclosureRelationshipStrength2012},
and the perceived intimacy of the content~\cite{maAnonymityIntimacySelfDisclosure2016}.
This echoes other concerns about sharing frisson data with complete strangers (P35)
and frisson experiences as privacy (P34).

On the other side of the coin are positive comments
like the non-identifiability of aggregated data (P7, P29) and
the prevention of uncomfortable messages in non-text-based communication (P10).
Self-disclosure is also influenced by the perceived usefulness of the data~\cite{schneegassUnderstandingImpactInformation2019}
and is sometimes used strategically to reach social goals~\cite{omarzuDisclosureDecisionModel2000}.
We can also identify similar rationales among those who are willing to share their data,
like to become ``a part of the group'' (P37)
and to bring stronger affective experiences to other viewers (P7, P43).
Such debates are more present on the sender's side,
which truly reflect the salient social cue of this affective communication channel.
But it also calls for users' control over the sharing action,
like a toggle for sensing and sharing (P34),
a retrospective view of the user's own frissons prior to sharing (P24),
or specifying the targets to share with (P30, P39, P44, P46).

Given the overall social cue of this frisson-sharing feature
and users' mixed attitudes towards it,
we further delved into the rationales behind users' different mindsets
in the next section.
They eventually lead to a summary of design considerations
for future designers to ensure satisfactory user experiences.

\subsection{Design Considerations}\label{sec:dc}

\subsubsection{Customizing the Frisson Feedback Designs}

As discussed in Sections~\ref{sec:comm:use} and~\ref{sec:use},
users may have diverse opinions on the understandability and usability of different frisson feedback designs.
This phenomenon underscores the importance of providing different modalities and encodings
of the audience's frisson data
to suit different users' tastes.
We further identified three major purposes of using this frisson-sharing feature
from participants' comments,
which also echoes previous literature on multimedia-induced frissons~\cite{harrisonThrillsChillsFrissons2014}.
1) \textbf{To feel the data --- seeking resonance and the sense of ``watching together''}.
Some people experience strong psycho-physiological responses when they perceive
deep empathy or affective contagion~\cite{harrisonThrillsChillsFrissons2014}.
Similarly in our findings,
some participants are heartfeltly curious about other viewers' feelings
and open to establishing connections with others.
They may feel happy when resonance with other viewers occurs
(P7, P19, P22, P26, P28, P37, P42, P43, P44)
and also socially connected~\cite{inagakiSharedNeuralMechanisms2013}.
In this case,
the frisson feedback shall have an intuitive reference to the feeling of frissons (P43, P44).
2) \textbf{To use the data --- augmenting the video-watching experience}.
Some participants ultimately want to enhance their own user experience via
the additional stimuli.
This goal is consistent to another frisson mechanism: strong sensory stimuli~\cite{harrisonThrillsChillsFrissons2014}.
Those participants may be aware that the stimuli are generated from others' physiological reactions,
but they do not seek a profound social interpretation of the information.
To them, the frisson data may be of little difference from other data sources,
such as the volume, the rhythm, or handcrafted data by the video's author.
In this case,
the frisson feedback can be more intense,
creating a stronger sensory impression.
Previous works of
thermo-haptic stimulus~\cite{heFrissonWavesExploring2022,jainFrissonLeveragingMetasomatic2022,haarAugmentingAestheticChills2020},
shall be also useful here because they intended to synthesize additional frissons.
3) \textbf{To analyze the data --- overviewing the audience's reactions}.
Some participants like to put on an analytical mindset,
and they are especially interested in seeing the overall real-time affective states
of the audience members.
They want to use this data to navigate to parts of the content that are aesthetically arousing or to understand the aesthetic values of the video by paying more attention to other people's frisson moments.
In the long run, such information may enhance their own frissons
induced by aesthetic appraisal~\cite{harrisonThrillsChillsFrissons2014}.
In this case,
designers can provide more analytical, dashboard-like visualizations
of the data (P32, P40, P45)
with evident marks of ``how large is the largest'' (P32, P40).

Note that even under the same purpose,
users may still have divergent preferences for the modalities and encodings of the frisson feedback
due to aesthetic preferences,
the attention demands,
and/or the current balance
between social needs
and immersion in the video.
Whether the feedback suits their preferences may also influence their perception of
the social distance with other viewers and their own affective experience of watching
online videos (P15, P20, P42, P44).
Therefore,
designers should also allow an appropriate degree of feature customization,
so that users can find their favorite way to consume the data.

\subsubsection{Mitigating Potential Negative Impacts on Affective Experience}

Despite the improvement in the efficiency of affective communication,
there are moments when users' affective experiences are adversely influenced,
especially when the users' own moments of frissons or emotional arousal
coincide with troughs of the data,
or vice versa.
For users that \textbf{feel the data},
it may trigger negative emotions of
being not understood, confused, or self-skeptical (see Section~\ref{sec:comm:affect}).
For users that \textbf{use the data},
it may also trigger negative user experiences like
interruption of their arousal
and false expectations of the functionality (see Section~\ref{sec:comm:use}).
Therefore,
it can be plausible to avoid such negative impacts by
only showing the data peaks when a sufficiently large portion of the audience have frissons (P25),
or only when the peaks match the user's own frissons (which are detected in real time) (P46).

However,
many participants were \textit{not} bothered by such ``mismatching'' moments,
and some especially praised the authenticity of the feedback (P10, P42, P47).
And for users who \textbf{analyze the data},
it is also essential to reflect the real data.
Hence,
designers may allow users to control the aforementioned feature.

\subsubsection{Privacy Concerns}

Although biosignal data
and their derived affective information
is sometimes considered private~\cite{liuCanBiosignalsBe2017,liuSupportingSocialInteractions2017},
few participants expressed privacy concerns regarding this new affective communication channel in our study because we maintained full data transparency and user agency in the process.
However,
they laid out two specific requirements
about contributing their frisson data on real-world online video platforms.
First, strict data management regulations are desired on
whatever party that would host the frisson-sharing service
(e.g., the online video platform itself or a third-party platform that adds to it).
It must refrain from monetizing users' physiological data (P30, P39) or storing it with personal identity attached (P26, P27).
Also, it should inform of what data is collected (P1, P46).
Second,
participants also require ``social privacy'' ---
controlling when and to whom they want to share their frisson data
to prevent negative social consequences.
This is elaborated in Sections~\ref{sec:willingness} and \ref{sec:social-sharing}.
Our current design provides such affordances via
complete user control of both the physical sensing devices
and the on/off of frisson sharing on video platforms.
It is also suggested that users can review the recorded data before publishing it (P24).

\subsection{Generalizability}

\subsubsection{Generalizability to Other Scenarios}\label{sec:gen}

In this paper,
we position sharing frissons on online video platforms
as an effortless, asynchronous, and auxiliary affective communication, which
accompanies individual experiences of media consumption.
As such,
it can be viewed as an alternative way (compared to conventional methods, e.g., comments) to engage with other users of online platforms 
and can be generalized to a broad range of usage scenarios in this regard.
First, our designs can be easily transferred to other (asynchronous and collective) online media appreciation activities such as listening to music, reading books, and playing video games on digital platforms.
Second, apart from sharing with a mass crowd,
another use scenario can be remote intimate sharing~\cite{wernerUnitedpulseFeelingYour2008,neidlingerAWElectricThatGave2017} to maintain close relationships.
As is also mentioned by our user study participants,
such a frisson-sharing feature,
whether asynchronous or synchronous,
promotes an ``accompanied watching'' experience (P26, P30),
and some of them expressed particular interest in
sharing with friends, family members, or romantic partners
(P30, P34, P39, P44, P46).
Third,
the auxiliary nature of our designs also inspired
some participants to imagine sharing
physiological responses to multimedia content on social media
such as X (formerly Twitter) and Instagram (P1, P17, P19).
Similar to sharing photos on social media~\cite{marlowPhotoSharingDiverse2011},
this shall enable richer non-collective and retrospective
annotations and narrations of the sharers' past experiences
and have more profound social effects.
Fourth,
the sensor-based and effortless affective sharing
is also beneficial to mixed reality and metaverse applications,
because the automatic detection fosters a more immersive user experience~\cite{paiPhysiologicalSignalDrivenVirtual2016}
and the affective communication channel
makes up for the missing social cues induced from physiological information~\cite{mogeSharedUserInterfaces2022}.

\subsubsection{Commercial Feasibility}\label{sec:feasibility}

We initially planned to use commercial smart wearable devices for
biosignal sensing and haptic feedback,
because it could best simulate the possible real-world deployment of the frisson-sharing feature
and eliminate the potential discomfort of wearing tailor-made sensors.
However, although EDA sensors have appeared in recent generations of commercial devices
such as FitBit trackers~\cite{fitbitinc.StressManagementFitbit},
relevant APIs for EDA data acquisition are yet to be opened to general developers.
We thus used customized sensing and vibration modules instead for this study while retaining a boundary of our design space according to the feasibility of implementing the features on existing smart wearables.
In spite of this,
the existence of EDA sensors on smart wearables
indicates that real-time frisson sensing
is technically applicable.
The sensor data, together with the vibration control,
can be communicated via Bluetooth.
Plus, the devices' battery life
will not be significantly affected,
as EDA sensors do not consume much power~\cite{luharukaMicrocontrollerbasedDataLogger2002},
and the sensors are only turned on during the period of an online video.
Regarding the haptic output module, we decided to implement the vibration feedback
on users' fingertips in our study
due to participants' concerns about mistaking vibration feedback
for push notifications.
To realize frisson-related vibration feedback on smart watches,
designers should try to distinguish the
vibrotactile patterns from push notifications~\cite{schneiderHapTurkCrowdsourcingAffective2016}.

\subsection{Future Research Directions}

In this section,
we propose several future research directions
for frisson-based affective communication.

Concerning the impact of sharing affective signals,
it would be interesting to see whether and how
users' own physiological responses to the video content
are influenced by other viewers' frisson data.
It would provide a more comprehensive view of physiological synchrony and affective contagion,
and whether the actual effect differs across designs.
We did not conduct this analysis
in our within-subject study because
it would have involved
watching the same video multiple times
(with frisson sharing on and off), which
would be likely to
introduce aesthetic fatigue
and influenced the result~\cite{robertsMixedmethodsExaminationAutonomous2020}.
Therefore,
we returned to self-reported physiological and emotional changes in the interview.
Nevertheless, such analyses may be possible in larger-scale, between-subject field studies deployed on commercial smart wearables (see Section~\ref{sec:feasibility}).
Users' biosignal data may also reveal
potential emotion saturation after experiencing too many frissons as well as whether the size of the audience has an impact on their experience.

Another topic is
whether users can identify the authenticity of the frisson data.
Previous work has found that sharing forged heart rates
in collaborative virtual environments
can influence users' perception of others' affective states~\cite{deySharingManipulatedHeart2019}.
It is interesting to see if viewers ---
especially those who \textbf{feel the data} ---
are affected by handcrafted fake frisson data,
in terms of their own experiences and the perceived social relationship with others.
From the senders' perspective,
sometimes people do not want to ``be authentic'' online~\cite{haimsonOnlineAuthenticityParadox2021};
therefore, researchers can investigate whether certain design decisions
of this communication channel may lead users to fake frisson data.

From the perspective of online video platforms,
the additional communication channel and the augmented user experience
may also alter users' browsing habits.
The acquisition of empathetic or sensory satisfaction may result in
a sensation seeking process~\cite{robertiReviewBehavioralBiological2004}
and an inclination for frisson-provoking videos.
It is left for future investigations
whether the viewer community will form a positive trend of
video commenting culture,
or it will negatively homogenize the content on online video platforms
into a concentration of sensory stimulation.

\subsection{Limitation and Future Work}\label{sec:limit}

There are still some limitations in this study.
In our design process and the user study, we mainly targeted young adults (aged 18--30).
Also,
the recruitment within a local university limited the coverage of
participants' cultural backgrounds and regional user habits of online video platforms.
Further studies can extend to a broader range of users with different ages and cultural backgrounds.

We acknowledge that
several inevitable subjective design decisions were made during the design process
and the choice of online videos.
For the design,
we tried to condense ideas and thoughts into the most appropriate designs
via a multi-stage design process,
but participants' preferences may still be biased by their past experiences, cultural backgrounds,
and other personal factors.
For the choice of online videos,
assuming that the frisson information is most useful
for arousing and frisson-triggering videos,
we selected videos based on previous literature and researchers' cross-validation.
Although it is out of this paper's scope to discuss which genres of videos
trigger the most episodes of frissons or are the most suitable for these frisson-sharing designs,
we are aware of the possibility of leaving out some videos where
users may be interested in sharing their frissons while watching in our study.
Therefore, we encourage future exploration of affective frisson feedback design
on different modalities and encodings,
with more genres of videos and for users from diverse cultural backgrounds.

\section{Conclusion}

In this paper,
we propose an affective communication approach on online video platforms
that senses, aggregates, and presents viewers' frissons
while watching videos asynchronously.
We have derived three designs of presenting the audience's frisson data
with minimal attention requirement
from a set of design activities,
and we evaluated them in a within-subject user study with 48 participants.

Through this research,
we verified the capability of frisson sharing as
an alternative and supplementary way to affectively engage with other users
and convey lightweight social cues.
Our deployment on online video platforms demonstrated the use of this affective communication channel
as ``spontaneous'' reactions when users engage in the same activity at different spaces and times.
We also attained a deeper understanding of how people
perceive this new supplementary, non-obtrusive, and authentic communication channel,
and we derived several design considerations accordingly.
We encouraged future investigation into such asynchronous, auxiliary, and affective biosignal sharing
for other scenarios,
as well as exploring affective use of physiological information in extended formats and its potential impact on media consumption.
\begin{acks}
    \edit{This research project is funded by the Seed fund of the Big Data for Bio-Intelligence Laboratory (Z0428) from The Hong Kong University of Science and Technology.}
\end{acks}

\bibliographystyle{ACM-Reference-Format}
\bibliography{main}



\end{document}